\documentclass[10pt,aps,amsmath,amssymb,prd,twocolumn]{revtex4-2}
 
\usepackage{bm} 
\usepackage[shortlabels]{enumitem} 
\usepackage{verbatim}
\usepackage{soul} 
\usepackage{graphicx}
\usepackage[T1]{fontenc}
\usepackage{xcolor,units}
\usepackage{hyperref}
\usepackage{listings}
\usepackage{tablefootnote}
\usepackage{xspace}
\usepackage[normalem]{ulem}
\hypersetup{
    unicode=false,          
    pdftoolbar=true,        
    pdfmenubar=true,        
    pdffitwindow=false,     
    pdfstartview={FitH},    
    pdfauthor={Boris Goncharov},     
    colorlinks=true,       
    linkcolor=blue,          
    citecolor=red,        
}
\graphicspath{{figures/}}

\newcommand{\dd}{{\rm d}}

\begin{document}

\definecolor{dkgreen}{rgb}{0,0.6,0}
\definecolor{gray}{rgb}{0.5,0.5,0.5}
\definecolor{mauve}{rgb}{0.58,0,0.82}

\lstset{frame=tb,
  	language=Matlab,
  	aboveskip=3mm,
  	belowskip=3mm,
  	showstringspaces=false,
  	columns=flexible,
  	basicstyle={\small\ttfamily},
  	numbers=none,
  	numberstyle=\tiny\color{gray},
 	keywordstyle=\color{blue},
	commentstyle=\color{dkgreen},
  	stringstyle=\color{mauve},
  	breaklines=true,
  	breakatwhitespace=true
  	tabsize=3
}

\newcommand{\mari}[1]{{\color{red}{MB: \textbf\small{#1}}}}
\newcommand{\IMRD}{\texttt{IMRPhenomD}\xspace}
\newcommand{\TAYF}{\texttt{TaylorF2}\xspace}
\newcommand{\IMRX}{\texttt{IMRPhenomXPHM}\xspace}

\title{\textsc{gwfish}: A simulation software to evaluate parameter-estimation capabilities of gravitational-wave detector networks}

\author{Ulyana Dupletsa$^{1,2}$}
\author{Jan Harms$^{1,2}$}
\author{Biswajit Banerjee$^{1,2}$}
\author{Marica Branchesi$^{1,2}$}
\author{Boris Goncharov$^{1,2}$}
\author{Andrea Maselli$^{1,2}$}
\author{Ana Carolina Silva Oliveira$^3$}
\author{Samuele Ronchini$^{1,2}$}
\author{Jacopo Tissino$^{1,2}$}
\affiliation{$^{1}$Gran Sasso Science Institute (GSSI), I-67100 L'Aquila, Italy}
\affiliation{$^{2}$INFN, Laboratori Nazionali del Gran Sasso, I-67100 Assergi, Italy}
\affiliation{$^3$Department of Physics, Columbia University in the City of New York, New York, NY 10027, USA}

\date{\today}

\begin{abstract}
An important step in the planning of future gravitational-wave (GW) detectors and of the networks they will form is the estimation of their detection and parameter-estimation capabilities, which is the basis of science-case studies. Several future GW detectors have been proposed or are under development, which might also operate and observe in parallel. These detectors include terrestrial, lunar, and space-borne detectors. In this paper, we present \textsc{gwfish}\footnote{\href{https://github.com/janosch314/GWFish}{github.com/janosch314/GWFish}}, a new software to simulate GW detector networks and to calculate measurement uncertainties based on the Fisher-matrix approximation. \textsc{gwfish} models the impact of detector motion on PE and makes it possible to analyze multiband scenarios, i.e., observation of a GW signal by different detectors in different frequency bands. We showcase a few examples for the Einstein Telescope (ET) including the sky-localization of binary neutron stars, and ET's capability to measure the polarization of GWs.
\end{abstract}

\maketitle

\section{Introduction}
The gravitational-wave (GW) community is currently in the phase of developing science cases and analysis tools of future GW detector networks \cite{MaEA2020,KaEA2021} including potential upgrades of the current infrastructures Virgo \cite{AcEA2015}, LIGO \cite{LSC2015}, LIGO India \cite{Sou2016}, KAGRA \cite{AkEA2018}, and the proposed Einstein Telescope \cite{PuEA2010,ET2020} and Cosmic Explorer \cite{EvEA2021}. The approved space-borne detector LISA is expected to begin observations in the second half of the 2030s \cite{ASEA2017}. Entirely new detector concepts are under study like the Lunar GW Antenna (LGWA) \cite{HaEA2021}. The networks can be formed between alike detectors observing at the same frequencies, but also combining observations in different frequency bands of the same GW sources \cite{Ses2016,Vit2016,GrHa2020}. Providing a simulation of conceivable observation scenarios formed by these detectors is an important part of the science-case development of future GW detector networks, but also a challenging and potentially computationally expensive task. In recent years, a few so-called Fisher-matrix codes were developed for this purpose \cite{ChEA2018,GrHa2020,Bor2021}.

In this article, we present the simulation software \textsc{gwfish}, which uses frequency-domain GW models combined with a time-domain simulation of GW detector networks through the stationary-phase approximation. This creates a framework to study important aspects of PE with future detector networks, where the change of position and orientation of a detector during the observation of a signal can have an important impact especially on the sky localization \cite{Cut1998,ChEA2018,GrHa2020,NiCa2021}. The calculation of PE errors is done using the Fisher-matrix approximation \cite{Val2008}, which corresponds to a Gaussian approximation of the likelihood expected to be acceptable for signals with higher signal-to-noise ratio. When the Fisher matrix is used to directly evaluate PE errors like in \textsc{gwfish}, then priors cannot be considered. As a consequence, overestimation and underestimation of PE errors are possible \cite{RoEA2013}. It should be noted that Gaussian likelihood approximations based on Fisher matrices can also be used for posterior sampling, which leads to a significant speed-up of the likelihood evaluation. This method can be combined with arbitrary priors.

A feature of \textsc{gwfish} is that multiband scenarios can be simulated as well. It is a simple step to do with Fisher-matrix codes since Fisher matrices for different detectors are added irrespective of the frequency band that provided the signal information. Technically, it just requires to set up the code so that detector motion and signal waveforms can be simulated accurately in different frequency bands. This includes the application of the time-delay interferometry (TDI) formalism to simulate space-borne detectors like LISA beyond the long-wavelength regime \cite{AET1999,PrEA2002,Val2005a}, and the support of multiple components of a detector, e.g., multiple interferometers of a xylophone configuration \cite{HiEA2010} or multiple sensors for LGWA type detectors \cite{HaEA2021}. Each component can be assigned a duty cycle for more realistic assessments of the observing scenario. Details of the detector simulation are described in section \ref{sec:sim}.

The Fisher-matrix formalism is computationally efficient, but numerically less robust than posterior sampling. The numerical challenges concern the calculation of waveform derivatives and the inversion of Fisher matrices. Waveform derivatives are best carried out in a hybrid analytical-numerical scheme to minimize the computational effort and numerical errors, and to keep it compatible with arbitrary waveform models (also referred to as waveform approximants). In \textsc{gwfish}, the numerical differentiation is tuned to the properties of waveform models to reduce numerical errors. Concerning the inversion of Fisher matrices, the main issue is that these matrices can be very close to singular, i.e., with a huge range of possible eigenvalues, which is connected to signal-model degeneracies. To get accurate PE errors at least of the parameters not involved in the model degeneracy, one needs to systematically deal with matrix singularity. The general PE framework of \textsc{gwfish} and how it addresses the numerical challenges are described in section \ref{sec:PE}.

As a demonstration, we show basic PE results for Einstein Telescope (ET) in section \ref{sec:examples}, which is a good example to understand the capabilities of \textsc{gwfish}, but also its limitations since single-detector scenarios are especially difficult to model in the Fisher-matrix formalism due to unavoidable degeneracies in the signal model. As a direct scientific conclusion of these analyses, we confirm that the ET configuration has unique capabilities for PE of compact-binary coalescences. Other more detailed studies carried out with \textsc{gwfish} have been published in separate papers. These include the evaluation of the perspectives of multi-messenger observations for ET (as single observatory and in network of GW detectors) operating in synergy with $\gamma$ and X-ray satellites using astrophysically motivated populations of BNS mergers \citep{Ronchini2022}, and very high-energy observatories (Banerjee et al. 2022 in preparation). \textsc{gwfish} was also used for the simulation of an astrophysical background in the ET null-stream study~\citep{goncharov2022null}.

\section{Detector-network simulation}
\label{sec:sim}
The simulation of a detector network needs to consider several aspects, which are represented under different categories inside \textsc{gwfish}:
\begin{itemize}
    \item Component (laser interferometer, TDI channel, accelerometer): time-varying detector response to GWs, noise modeling, duty cycle;
    \item Detector: one or multiple components observing the same frequency band and added as a group to a detector network;
    \item Network: one or multiple detectors observing together.
\end{itemize}
Most of the effort of the detector simulation goes into the calculation of the detector response. We outline the basics and refer to other publications for details. The long-wavelength approximation is applied to all terrestrial GW detectors simulated with \textsc{gwfish}, which assumes that the arm-length of a GW detector is much smaller than the length of a GW. It is not a fully accurate approximation above a kilohertz for a 40\,km detector like the proposed Cosmic Explorer \cite{Sch1997,AbEA2017a,EVE2017}. The response formalism for space-borne detectors can be adapted to ground-based detectors, e.g., to calculate the GW response of a Fabry-Perot cavity beyond the long-wavelength regime \cite{Sch1997}. 

The response of a detector is complicated by the fact that the detector orientation and position can change with time. Especially the change in orientation is important to simulate. Since signals are represented in frequency domain in \textsc{gwfish}, the question is how to convolve the detector response with the signal. In \textsc{gwfish}, it is assumed that a function $t(f)$ exists mapping signal frequencies to times. To obtain this function, we exploit the stationary phase approximation, which underlies most analytical models of compact-binary signals and is known to lead to very accurate approximations of the Fourier spectrum of compact-binary signals \cite{DrEA1999}. This function can be calculated from the phase $\phi(f)$ of the waveform:
\begin{equation}
    t(f)=\frac{1}{2\pi}\frac{\dd \phi(f)}{\dd f}.
\end{equation}
A technical issue with this approach to calculate $t(f)$ is that it requires a dense sampling in frequency space to have enough resolution of the phase $\phi(f)$ for a finite-difference method. Since we do not want to impose any unnecessary constraints on waveform modeling in \textsc{gwfish}, we instead calculate $t(f)$ directly from the lowest-order term of the phase evolution, e.g., see equation (15) in \cite{HaEA2013}. This approximation is acceptable since only the early phase of the inspiral needs to be modeled accurately to account for the effects of detector motion on PE. Once $t(f)$ is calculated for a signal, it can be used as a time vector for the simulation of the motion of the detectors. The GW signal of a detector can then be written in terms of a response tensor $\mathcal A(t(f))$ contracted with the GW strain tensor $h(f)$ \cite{SaSc2009}
\begin{equation}
    \mathcal R(f)=\mathcal A(t(f)):h(f).
\label{eq:resp}
\end{equation}
For example, the response tensor for a laser interferometer with two arms along direction $\vec e_1,\,\vec e_2$ assumes the form
\begin{equation}
    \mathcal A=\frac{1}{2}(\vec e_1\otimes\vec e_1-\vec e_2\otimes\vec e_2),
\end{equation}
where all quantities are functions of $t(f)$. It should be noted that in the case of terrestrial detectors, the actual frequency-dependent response of a GW detector, e.g., any aspect going into detector calibration, is included in the instrument-noise spectrum referred to GW strain. 
\begin{figure}[ht!]
\includegraphics[width=\columnwidth]{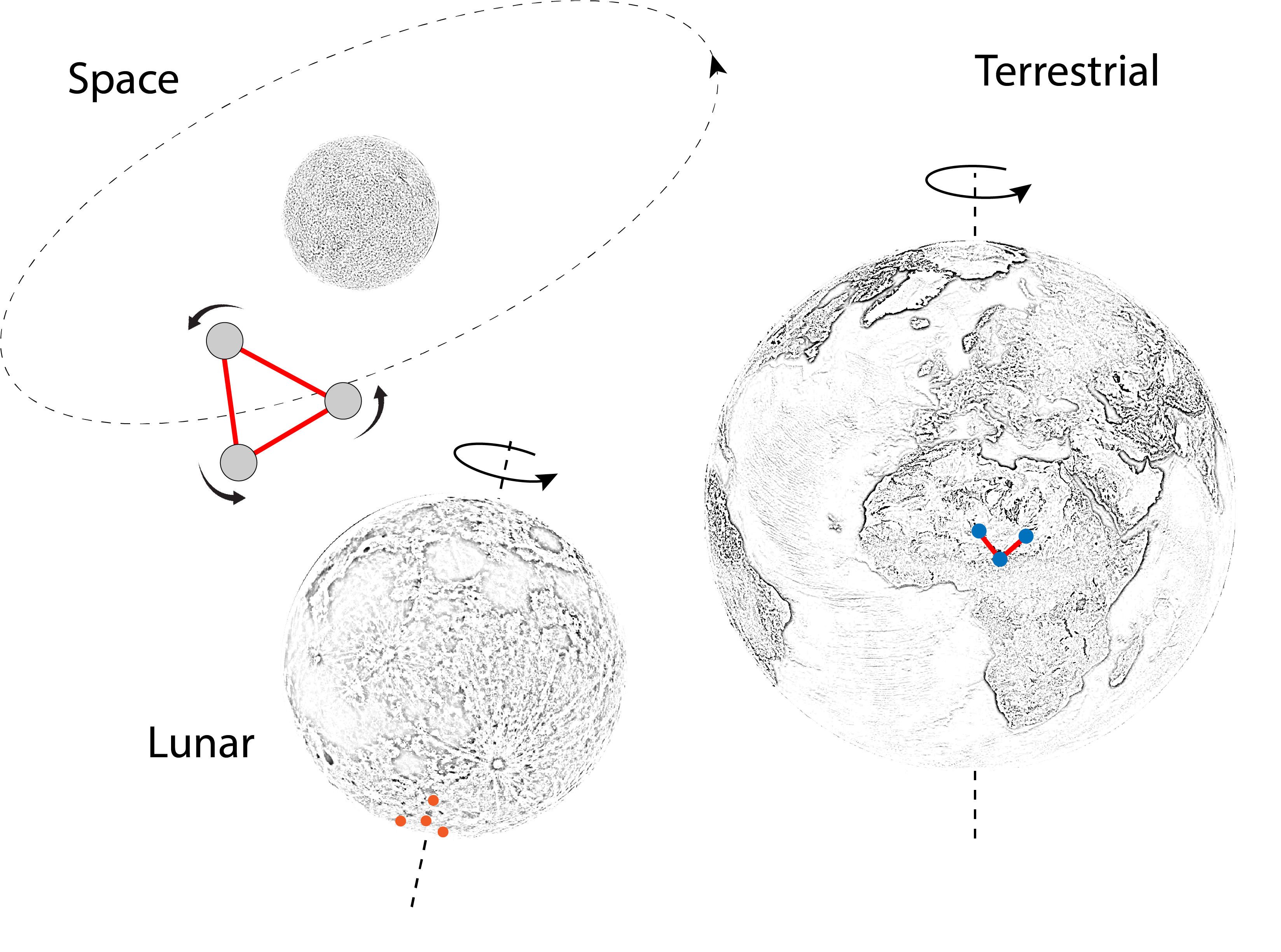}
\caption{Detectors included in \textsc{gwfish} with simulated detector motions indicated by arrows.}
\end{figure}

For LISA-type space detectors, another method must be used to calculate the GW response since the long-wavelength approximation does not hold \cite{AET1999}. The fundamental building blocks are the readouts $y_{ij}(f)$ of individual spacecraft links $j\rightarrow i$, which are then electronically combined with time delays to mimic the workings of a laser interferometer. The main purpose of combining different readouts is to reduce laser-frequency noise by several orders of magnitude, which is achieved in terrestrial detectors by optical interference. This method is called time-delay interferometry (TDI). Effectively, TDI makes each satellite of a triangular detector configuration a vertex of a laser interferometer. It turns out that the instrument noise in the basic Michelson-type TDI combinations is correlated, which means that it is convenient to diagonalize the noise-correlation matrix, which yields the three (noise) independent TDI channels $A,\,E,\,T$ named after the three scientists who set up the TDI formalism for LISA (Armstrong, Estabrook, and Tinto). The $T$ channel has a strongly suppressed response to GWs below $f_0=c/(2L)$, where $L$ is the average arm length of the space detector. The channel $T$ is the sum of the three Michelson-type TDI channels, and it is known as null stream in terrestrial GW detectors \cite{GuTi1989,ReEA2012}. The three independent channels $A,\,E,\,T$ are simulated in \textsc{gwfish} for space detectors. An approximation made in the calculation is that the so-called breathing motion of the space-detector configuration, i.e., a small modulation of the arm lengths over the course of a year, is neglected. Correcting the TDI combination for this effect is important for the suppression of laser-frequency noise, but it has a small effect on signal detection and PE. In other words, the satellite configuration is simulated as rigid body orbiting the Sun and with a rotating detector plane known as cartwheel motion \cite{Val2005a}.

The third detector type simulated in \textsc{gwfish} is LGWA, which consists of an array of seismometers measuring the surface displacement of the Moon produced by a passing GW. Here, the long-wavelength approximation applies again since the detector is designed for measurements in the decihertz band and below, and the length of the GWs need to be compared with the diameter of the Moon. Therefore, the response of a seismometer to GWs can again be described by equation (\ref{eq:resp}), and the response tensor takes the form
\begin{equation}
    \mathcal A=\vec e_n\otimes\vec e_1,
\end{equation}
where $\vec e_1$ is the displacement direction measured by the seismometer, and $\vec e_n$ is the surface normal vector at the seismometer position. The time dependence of the response comes from the rotation of the Moon. The effect of its orbital motion around Earth is neglected.

\section{Parameter estimation}
\label{sec:PE}
The PE is simulated in \textsc{gwfish} through the Fisher-matrix formalism, which is based on a Gaussian approximation of the likelihood function:
\begin{equation}
    \mathcal L(\vec x)\propto \exp(-\vec x^{\,\intercal}\cdot \mathcal C^{-1}\cdot\vec x/2).
\end{equation}
The covariance matrix $\mathcal C$ is the inverse of the Fisher matrix $\mathcal F$, and its components can be calculated from the signal model $h(\vec x\,)$,
\begin{equation}
    \mathcal F_{ij}=\sum\limits_{k=1}^N\langle \partial_{x_i} h^k| \partial_{x_j} h^k\rangle,
    \label{eq:fisher}
\end{equation}
where $N$ is the number of components of the GW detector network, and the derivatives of the waveform are calculated with respect to the model parameters $\vec x=(x_i)$. The expression makes use of an inner product defined on the signal-model manifold as follows:
\begin{equation}
    \langle a|b\rangle=4\int\limits_0^\infty\dd f \frac{\Re(a(\vec x,f)b^*(\vec x,f))}{S_\text{n}(f)}.
\end{equation}
The product requires a model of the instrument noise spectral density $S_\text{n}(f)$. The theoretical framework is simple, but there are practical challenges. First, for some of the model parameters, the derivatives need to be calculated numerically. It is not necessary for all waveform models, but one ought to adopt an approach that is valid for all models. Numerical differentiation suffers from limitations of numerical precision, e.g., some derivatives might be calculated from tiny variations of potentially very large numbers. In \textsc{gwfish}, derivatives with respect to waveform phase at merger $\phi_\text{c}$, merger time $t_\text{c}$, and source luminosity distance $D$ are calculated analytically. Derivatives of all other parameters (masses, spins, tidal deformation, sky location, inclination angle, polarization angle) are calculated numerically. The waveform phase is a sum of terms with large and small values. Especially the component $2\pi f t_\text{c}$ can take very large numerical values, and \textsc{gwfish} eliminates this term in calculations of numerical derivatives involving the phase of the waveform.

Another numerical challenge is connected to the inversion of the Fisher matrix in equation (\ref{eq:fisher}) to obtain the PE errors. The first issue is that model parameters can have a large range of numerical values, e.g., $10^{30}$ for masses in kg and values of order 1 for all angular variables. The situation can be ameliorated by normalizing the Fisher matrix \cite{PMJ2016}:
\begin{enumerate}
    \item Collect the diagonal elements $d_i=\mathcal F_{ii}$ of the Fisher matrix.
    \item Divide row $k$ of the Fisher matrix by $\sqrt{d_k}$. Divide column $k$ of the Fisher matrix by $\sqrt{d_k}$. This leaves a matrix with diagonal components equal to one and all off-diagonal elements in the range $[-1,1]$.
    \item Invert the matrix, then repeat step (2) with the inverted matrix using the same values for $d_k$.
\end{enumerate}
A remaining potential numerical issue is when Fisher matrices are close to degenerate.
To address this issue, \textsc{gwfish} excludes all singular values lying below a certain numeric precision threshold from the matrix before inverting it \cite{PrEA2007}. The main outcome of this step is to make the matrix inversion robust with respect to all significant singular values containing physical information. In cases where components of small singular values need to be discarded, the waveform model is close to degenerate with respect to at least one parameter pair, e.g., distance and inclination angle can show strong correlations. The PE errors estimated for such strongly correlated parameters are typically substantially different from PE errors coming out of a full Bayesian analysis (posterior sampling). For example, usage of prior distributions in Bayesian analyses can substantially improve parameter estimates in these cases. Model degeneracy is one of the fundamental failure modes of the Fisher-matrix analysis.  

\textsc{gwfish} can use all waveform models of LALSimulation \cite{lalsuite}, and it provides independent implementations of the \TAYF (of 3.5th post-Newtonian order) and \IMRD approximants. It is beyond the scope of this paper to provide a useful guideline for approximant usage with \textsc{gwfish}. Generally, it is important to understand the physics that the LALSimulation approximant incorporate, and these can have an enormous impact on parameter estimation. For example, it should be clear that for the analysis of high-mass or high-redshift BBH signals, which appear only with several cycles in the observation band, an approximant like \TAYF is inappropriate since it does not include the merger and ringdown of the binary. In these cases, the safest choice is to use approximants that include higher-order modes. However, there are good use cases even for \TAYF, e.g., if one wants to calculate signal-to-noise ratio (SNR) and localization of binary neutron stars. It is recommended to make comparison of results for a substantial random sample of signals (of order 10 -- 100 signals) before deciding, whether it is safe to use a simplified waveform model.

\subsection{Einstein Telescope}
The Einstein Telescope (ET) is a proposed underground infrastructure in Europe to host future-generations of GW detectors. With respect to its PE capabilities, three features stand out, which will make it a unique facility among future detectors:
\begin{itemize}
    \item It has the geometry of an equilateral triangle and laser-interferometer arms launching from each of its vertices. It is expected that this triangular configuration will give it unique capabilities concerning the measurement of GW polarizations and parameters correlated with GW polarizations.
    \item It will have 10\,km arm length, which, together with several technology improvements with respect to current detectors, leads to a greatly improved peak sensitivity by about a factor 10. Binaries of compact objects will be detected along the cosmic history; BNS and BBH mergers are both expected to be detected with a rate of $10^5$ per year. The enormous increase of SNR for close events will enable accurate parameter estimation, enabling great precision in GW astronomy.
    \item It will consist of a total of 3 pairs of laser interferometers. Each pair includes a low-frequency interferometer and a high-frequency interferometer together covering an observation band from 3\,Hz to several kHz. The observation is thus extended to much lower frequencies compared to current GW detectors, which leads to longer observation times of individual compact-binary signals, and it will also make it possible to observe larger-mass compact binaries well into the regime of intermediate-mass black-hole binaries. Long observation times enable stand-alone sky-location measurements by ET of neutron-star binaries exploiting amplitude modulations due to Earth's rotation. Due to the fact that the observed mass is the redshifted source-frame mass, i.e., multiplication by $(1+z)$, the low-frequency sensitivity will make it possible to access BBH mergers in the early Universe where primordial BHs can be discriminated from BHs with stellar origin \citep[e.g.][]{Ng2021,DeLuca2021}.
\end{itemize}

\subsection{\textsc{bilby} versus \textsc{gwfish}}
We  compare SNRs and PE between \textsc{bilby} --- a detector simulation and Bayesian analysis software \cite{AsEA2019} --- and \textsc{gwfish} for four signals observed with ET (see Table \ref{tab:signal_sum}). Due to current constraints imposed by \textsc{bilby}, and by other state-of-the-art PE software, it is not possible to carry out accurate Bayesian analysis of BNS signals for ET. This has to do with computational limitations. Standard representations of BNS waveforms for ET would require about 10$^8$ samples due to the day-long observation time of these signals with ET, and \textsc{bilby} also does not consider Earth's rotation during the observation of GW signals for waveform modeling (an effect that does not play a role for analyses of compact binaries with current GW detectors). As a consequence, we carry out our comparison for BBHs, which, in some sense, is also more interesting since model degeneracies play a more important role. We do the comparison for each signal with a random realization of the detector noise in \textsc{bilby} in order to compare the \textsc{gwfish} results with a more realistic simulation of the Bayesian analysis. However, while noise realizations can generally influence PE errors \cite{Val2011}, it turns out that it has a negligible effect here since the SNRs of all signals are very high.

Signals 1 -- 3 have randomly sampled angular parameters, while the fourth signal (Signal 3*) has the same parameters as Signal 3 except for the two masses which are the same as for Signal 1. The signals are at relatively low redshift to make sure that the SNRs are high. In this case, the Gaussian approximation of the likelihood function is expected to hold, but we will see, not entirely surprisingly, that PE errors produced by \textsc{gwfish} and \textsc{bilby} are still not identical. As for the detector, we employ ET alone with its three components. For both \textsc{bilby} and \textsc{gwfish} we are using the inspiral-merger-ringdown waveform \IMRD from LALSimulation, setting the minimum frequency to 10\,Hz and the maximum frequency to 1024\,Hz. For \textsc{gwfish}, we include results with 2\,Hz lower bound to show the impact of ET's low-frequency sensitivity on PE. The reason not to use a 2\,Hz lower bound for waveforms in \textsc{bilby} is that \textsc{bilby} enforces a frequency resolution of about 1/(duration of signal), which means that the number of samples used for each waveform becomes very large even for BBHs with a 2\,Hz lower bound compared to 10\,Hz. Consequently, it takes too long to finish posterior sampling. In our analysis, we used a frequency resolution of 1/153\,Hz for Signal 1, 2 and 3, and 1/108\,Hz for Signal 3*. A simpler waveform model such as \TAYF would have been inappropriate here since model degeneracies would be higher especially when simulating BBH observations involving only ET. 
\begin{table}[ht!]
\centering
\caption{Parameters\footnote{{\bf For precise comparisons with more significant digits, one can access the injection files on the \textsc{gwfish} \texttt{GitHub} repository under the folder \texttt{injections/injections$\_$paper} together with a Jupyter notebook, \texttt{gwfish$\_$paper1$\_$table3.ipynb}, to reproduce the results.}} of the signals used for SNR and PE comparisons between \textsc{bilby} and \textsc{gwfish}. The spins are assumed to be aligned with the orbital angular momentum.}
{\small
\begin{tabular}{r|lll}
\toprule
            	&\textbf{Signal 1}		&\textbf{Signal 2}		&\textbf{Signal 3, 3*}\\
	\hline
	\hline
    $\text{RA}$ [rad]		&$0.375$		        &$3.026 $		    &$4.154$\\   
    \hline
    $\text{DEC}$ [rad]		&$-0.363$	            &$-0.282$		    &$-0.972$ \\   
     \hline
    $\Psi$ [rad]			&$1.738$	            &$4.552$			&$2.071$\\   
    \hline
    $\iota$ [rad]			&$1.645$                &$1.143$		  &$1.261$\\   
    \hline
    $d_{\rm{L}}$ [Mpc]	    &$453$	                &$453$		        &$453$\\
     \hline
    $z$						&$0.098$	            &$0.098$			&$0.098$\\
    \hline
    $m_{1, \rm src} [M_{\odot}]$	&$10.0$	    &$30.0$	        & 60.0, 10*\\   
    \hline
    $m_{2,\rm src} [M_{\odot}]$	&$5.0$	    &$15.0$		        &30.0, 5*\\ 
    \hline
    $t_{\rm c}$ [GPS]	&$1120381489.604$	&$1124867056.193$ &$1122192865.797$\\   
    \hline
    $\text{phase}$				&$3.472$		    &$1.098$			&$3.441$\\
    \hline
    spin $a_1$				&$0.729$		    &$0.729$			&$0.729$\\
    \hline
    spin $a_2$				&$0.620$		    &$0.620$			&$0.620$\\
    \hline
\end{tabular}
}
\label{tab:signal_sum}
\end{table}
SNR results for these signals are summarized in table \ref{tab:snr_cfr}. The SNRs match between \textsc{gwfish} and \textsc{bilby} with expected minor deviations, for example, because \textsc{gwfish} simulates the effects of the rotation of Earth on the waveform while \textsc{bilby} does not.

\begin{table}
\centering
\caption{The comparison between SNR for the four different signals (see Table~\ref{tab:signal_sum}) between \textsc{bilby} and \textsc{gwfish}. The \IMRD waveform model was used with $f_{\rm{min}}=10$\,Hz and $f_{\rm{max}}=1024$\,Hz. The SNRs reported here are the sum (in quadrature) over ET's three components. Note that the SNR calculation with \textsc{bilby} was done using the same waveform and the same instrument-noise spectrum as for \textsc{gwfish}. The main objective of the SNR comparison was to assess how well the two simulations match in terms of coordinate systems and detector models.}
{\small
\begin{tabular}{r|ll}
\multicolumn{3}{c}{\textbf{SNR}}\\
\hline
\hline
    \textbf{Signal}		&\textbf{\textsc{bilby}}     &\textbf{\textsc{gwfish}}\\
	\hline
	\hline
    Signal 1		&$72.1$		&$72.1$\\
	\hline
    Signal 2		&$219$		&$219$\\
	\hline
    Signal 3		&$1038$	    &$1038$\\
    \hline
    Signal 3*        &$236$       &$236$\\
\end{tabular}
}
\label{tab:snr_cfr}
\end{table}

For the comparison of PEs between \textsc{bilby} and \textsc{gwfish}, we use the same signals of table \ref{tab:signal_sum}. Posterior sampling requires to specify a prior on the parameters (at the very least a uniform prior to constrain the reasonable/physical range of parameter values). We used:
\begin{itemize}
    \item uniform priors for \textit{mass ratio} and \textit{chirp mass}
    \item uniform prior for the \textit{polarization angle} $\Psi$
    \item uniform prior on the merger rate in comoving volume (fixing the prior on the \textit{luminosity distance} $d_{\rm{L}}$)
    \item isotropic distribution in the sky (fixing the priors on \textit{right ascension RA} and \textit{declination DEC})
    \item uniform prior for the cosine of the \textit{inclination angle} $\iota$
\end{itemize}
The other 8 parameters (6 spin parameters, phase and time) are given a delta prior and therefore are not taken into consideration for posterior sampling and are excluded also from the Fisher matrices. \textsc{bilby} results are obtained using the \emph{PEsummary} package \cite{Hoy:2020vys}. In table \ref{tab:pe_summary}, we report the recovered median values for the parameters and the relative errors at the $90\%$ confidence level for \textsc{bilby}. The parameter values reported for \textsc{gwfish} are the true waveform parameters. \textsc{gwfish} outputs $1\sigma$ PE errors, which are converted to the $90\%$ confidence level in Table \ref{tab:pe_summary}, which can be done in Gaussian approximation by multiplying the errors by $1.645$. Moreover, under the \textsc{gwfish} results column, for each signal, we have two results per parameter: the upper one is obtained using $f_{\rm min}=10$\,Hz, the lower one with $f_{\rm min}=2$\,Hz. Starting from 2\,Hz yields smaller uncertainties on the parameters by about a factor 2. 

Generally, there is good agreement between \textsc{gwfish} and \textsc{bilby} PE errors when taking \textsc{gwfish} results with 10\,Hz low-frequency cutoff. For example, mass errors typically match within a factor 2. There are exceptions, which are all connected to multi-modal posterior distributions, e.g., several errors estimated with \textsc{bilby} are significantly larger since they include the contribution of a second mode, which the Gaussian likelihood approximation cannot reproduce. Interesting to point out is that at least for these 4 signals, which all have very high SNR, $\iota$, $d_{\rm L}$ are both estimated well. This tells us that ET, as suspected in the past due to its triangular configuration, is able to break the inclination-distance degeneracy. It needs to be noted though that our small and high-SNR sample of signals is not sufficient to draw general conclusions about ET PE capabilities, and errors might also increase when spins, merger phase and time are included in the analysis. What we can conclude from this table is that correlations between model parameters do not necessarily invalidate the \textsc{gwfish} PE errors, which means that Fisher-matrix based analyses can produce useful estimates of PE errors.

\begin{table*}
\centering
\caption{Comparing measurement uncertainties obtained with \textsc{bilby} and \textsc{gwfish}. We use the signals outlined in Table \ref{tab:signal_sum} and \IMRD as waveform model ($f_{\rm{min}}=10$\,Hz and $f_{\rm{max}}=1024$\,Hz). The errors are at $90\%$ confidence/credibility. The \textsc{gwfish} column presents two results for each parameter: the upper one is obtained starting from $f_{\rm{min}}=10$\,Hz, the one below is obtained starting from a minimum frequency of 2\,Hz. The parameter values reported for \textsc{gwfish} are the true waveform parameters, while the values reported for \textsc{bilby} are the median values of the posterior distribution.}
{\small
\begin{tabular}{r|ll|ll|ll|ll}
\toprule
&\multicolumn{2}{c|}{\textbf{Signal 1}} &\multicolumn{2}{c|}{\textbf{Signal 2}} &\multicolumn{2}{c|}{\textbf{Signal 3}} &\multicolumn{2}{c}{\textbf{Signal 3*}}\\
\hline
\hline
    Parameter   &\textsc{gwfish}   &\textsc{bilby}    &\textsc{gwfish}  &\textsc{bilby}   &\textsc{gwfish}  &\textsc{bilby} &\textsc{gwfish}  &\textsc{bilby}\\
     \hline
    $\text{RA}$ [rad]
    &\begin{tabular}{c}
    $0.37^{+0.3}_{-0.3}$    \\  $0.37^{+0.18}_{-0.18}$
    \end{tabular}
    &$2.2^{+3.5}_{-1.8}$
    &\begin{tabular}{c}
    $3.0^{0.004}_{-0.004}$   \\  $3.0^{+0.0031}_{-0.0031}$ 
    \end{tabular}
    &$1.2^{+1.9}_{-0.009}$
    &\begin{tabular}{c}
    $4.1^{+0.69}_{-0.69}$    \\   $4.1^{+0.62}_{-0.62}$
    \end{tabular}
    &$4.2^{+0.28}_{-0.11}$
    &\begin{tabular}{c}
    $4.1^{+1.7}_{-1.7}$    \\   $4.1^{+0.75}_{-0.75}$
    \end{tabular}
    &$4.6^{+0.45}_{-0.47}$\\
    \hline
    $\text{DEC}$ [rad]	
    &\begin{tabular}{c}
    $-0.36^{+0.19}_{-0.19}$     \\      $-0.36^{+0.12}_{-0.12}$
    \end{tabular}
    &$-0.3^{+1.1}_{-0.15}$ 
    &\begin{tabular}{c}
    $-0.28^{+0.0037}_{-0.0037}$   \\     $-0.28^{+0.0031}_{-0.0031}$
    \end{tabular}
    &$-0.82^{+0.54}_{-0.04}$    
    &\begin{tabular}{c}
    $-0.97^{+0.64}_{-0.64}$ \\ $-0.97^{+0.58}_{-0.58}$
    \end{tabular}
    &$-1.0^{+0.32}_{-0.087}$
    &\begin{tabular}{c}
    $-0.97^{+1.6}_{-1.6}$ \\ $-0.97^{+0.7}_{-0.7}$
    \end{tabular}
    &$-0.8^{+0.37}_{-0.29}$\\   
     \hline
    $\Psi$ [rad]
    &\begin{tabular}{c}
    $1.7^{+0.3}_{-0.3}$	    \\ $1.7^{+0.18}_{-0.18}$  
    \end{tabular}
    &$2.2^{+3.1}_{-1.8}$ 
    &\begin{tabular}{c}
    $4.5^{+0.015}_{-0.015}$      \\        $4.5^{+0.012}_{-0.012}$ 
    \end{tabular}
    &$1.4^{+2.7}_{-0.44}$ 
    &\begin{tabular}{c}
    $2.1^{+0.45}_{-0.45}$  \\     $2.1^{+0.41}_{-0.41}$
    \end{tabular}
    &$2.0^{+0.064}_{-0.21}$
    &\begin{tabular}{c}
    $2.1^{+1}_{-1}$  \\      $2.1^{+0.49}_{-0.49}$
    \end{tabular}
    &$4.8^{+0.46}_{-3.3}$\\
    \hline
    $\iota$ [rad]
    &\begin{tabular}{c}
    $1.6^{+0.12}_{-0.12}$   \\     $1.6^{+0.074}_{-0.074}$    
    \end{tabular}
    &$1.9^{+0.044}_{-0.045}$ 
    &\begin{tabular}{c}
    $1.1^{+0.039}_{-0.039}$      \\$1.1^{+0.031}_{-0.031}$
    \end{tabular}
    &$1.1^{+0.048}_{-0.06}$  
    &\begin{tabular}{c}
    $1.3^{+0.005}_{-0.005}$   \\ $1.3^{+0.004}_{-0.004}$
    \end{tabular}
    &$1.3^{+0.001}_{-0.002}$
    &\begin{tabular}{c}
    $1.3^{+0.013}_{-0.013}$   \\ $1.3^{+0.0071}_{-0.0071}$
    \end{tabular}
    &$1.3^{+0.006}_{-0.007}$\\   
    \hline
    $d_{\rm{L}}$ [Mpc]
    &\begin{tabular}{c}
    $453^{+390}_{-390}$	     \\$453^{+241}_{-241}$ 
    \end{tabular}
    &$576^{+70}_{-122}$
    &\begin{tabular}{c}
    $453^{+37}_{-37}$       \\$453^{+29}_{-29}$   
    \end{tabular}
    &$477^{+53}_{-44}$ 
    &\begin{tabular}{c}
    $453^{+6}_{-6}$  \\ $453^{+5}_{-5}$
    \end{tabular}
    &$453^{+1.1}_{-1.2}$
    &\begin{tabular}{c}
    $453^{+14}_{-14}$  \\ $453^{+6}_{-6}$ 
    \end{tabular}
    &$453^{+4.7}_{-4.6}$\\
    \hline
    $\text{m}_{1, \rm src} [M_{\odot}]$	
    &\begin{tabular}{c}
    $10^{+0.35}_{-0.35}$  \\	$10^{+0.19}_{-0.19}$
    \end{tabular}
    &$9.8^{+0.33}_{-0.38}$     
    &\begin{tabular}{c}
    $30^{+0.46}_{-0.46}$  \\$30^{+0.26}_{-0.26}$
    \end{tabular}
    &$30^{+0.62}_{-0.63}$   
    &\begin{tabular}{c}
    $60^{+0.13}_{-0.13}$\\$60^{+0.12}_{-0.12}$
    \end{tabular}
    &$60^{+0.095}_{-0.092}$
    &\begin{tabular}{c}
    $10^{+0.11}_{-0.11}$\\$10^{+0.061}_{-0.061}$
    \end{tabular}
    &$10^{+0.11}_{-0.1}$\\   
    \hline
    $\text{m}_{2,\rm src} [M_{\odot}]$
    &\begin{tabular}{c}
    $5.0^{+0.15}_{-0.15}$	  \\$5.0^{+0.083}_{-0.083}$ 
    \end{tabular}
    &$4.9^{+0.19}_{-0.16}$ 
    &\begin{tabular}{c}
    $15^{+0.2}_{-0.2}$  \\ $15^{+0.11}_{-0.11}$
    \end{tabular}
    &$15^{+0.17}_{-0.17}$   
    &\begin{tabular}{c}
    $30^{+0.06}_{-0.06}$\\$30^{+0.05}_{-0.05}$
    \end{tabular}
    &$30^{+0.04}_{-0.04}$
    &\begin{tabular}{c}
    $5.0^{+0.046}_{-0.046}$\\$5.0^{+0.026}_{-0.026}$
    \end{tabular}
    &$5.0^{+0.046}_{-0.045}$\\ 
    \hline
\end{tabular}
}
\label{tab:pe_summary}
\end{table*}

Corner plots of the posterior samples with \textsc{bilby} are shown in the appendix in Figures \ref{fig:signal1_corner} to \ref{fig:signal4_corner}. These also report the contours from \textsc{gwfish} estimates. We would like to highlight problems with the \textsc{bilby} parameter estimation for the low-mass system of Signal 1. This choice of parameters led to a multi-modal posterior distribution for the two masses. This happened with a rather broad prior on the mass values. We did not manage to tune the settings of the posterior sampling algorithm to correctly sample the posterior, e.g., by changing the number of sampling points or the correlation length. We then decided to reduce the width of the mass priors to a value just small enough to avoid a multi-modal distribution.

A Fisher matrix analysis is expected to break down when the signal SNR is low. Therefore, we show here a case for SNR=8, where once again we compare \textsc{bilby} to \textsc{gwfish} parameter-estimation errors. For this analysis, we used the parameters from Signal 3, but we increased its distance to obtain SNR=8. Since the observed masses now differ significantly from the intrinsic masses, we report the analysis for the two masses in the detector frame, instead of the source frame used in the previous analysis. We used a frequency range from $10$\,Hz to $1024$\,Hz, with a frequency resolution of $1/153$. Table \ref{tab:low_snr} reports the results at $90\%$ C.L.\;from both \textsc{bilby} and \textsc{gwfish}. In contrast to the high-SNR signals, \textsc{gwfish} does not accurately estimate the errors for sky location and polarization angles. The priors of the model parameters play a more important role in low-SNR observations. A comparison with the full posterior obtained from \textsc{bilby} is shown in the corner plot of Figure \ref{fig:signal3_low_snr_corner}.

\begin{table}
\caption{Comparison between \textsc{bilby} and \textsc{gwfish} for Signal 3 put at a distance such that SNR=8. The reported errors are at $90\%$ C.L.}
\centering
\begin{tabular}{r|c c}
\toprule
&\textsc{bilby} &\textsc{gwfish}\\
\hline
\hline
$\text{RA}$ [rad]
    &$4.5^{+0.5}_{-0.5}$ &$4.1\pm 83$\\
\hline
$\text{DEC}$ [rad]
    &$-0.66^{+0.32}_{-0.39}$ &$-0.97\pm 81$\\
\hline
$\text{$\Psi$}$ [rad]
    &$1.8^{+0.35}_{-0.43}$ &$2.1\pm 54$\\
\hline
$\text{$\iota$}$ [rad]
    &$1.2^{+0.19}_{-0.31}$ &$1.3\pm 0.6$\\
\hline
$\text{d$_{\rm{L}}$}$ [Mpc]
    &$68508^{+26173}_{-18288}$ &$58115\pm 64040$\\
\hline
$\text{m}_{1,\rm det}$ [$M_{\odot}$]
    &$61.3^{+7.6}_{-7.2}$ &$65.9\pm 18$\\
\hline
$\text{m}_{2,\rm det}$ [$M_{\odot}$]
    &$35.2^{+4.3}_{-3.6}$ &$32.9\pm 8$\\
\hline
\end{tabular}

\label{tab:low_snr}
\end{table}

\subsection{\TAYF versus \IMRD versus \IMRX}
\label{subsec:wfs}
Next, we compare results obtained with two approximants: \TAYF (see \cite{TF2_1} and \cite{TF2_2}) and \IMRD (see \cite{Phenom1} and \cite{Phenom2}). The \TAYF model is the simplest frequency-domain approximant, which describes the inspiral phase, and accurately only the early inspiral, which is sufficient for estimates, e.g., of the chirp mass and BNS localization through Earth's rotation. We apply a high-frequency cutoff at 4 times the frequency of the innermost stable circular orbit $f_{\rm isco}=6^{-1.5}\pi^{-1} c^3/(GM)$ to approximately match the SNRs obtained with other waveform models. The \IMRD model extends over the full inspiral-merger-ringdown of a compact-binary signal, but there are still constraints on the physics described by this model. For example, it only models the fundamental mode of the GW signal and systems with aligned spins. The purpose of this comparison is to show how a waveform model affects parameter estimation, even at the simple level of Fisher-matrix based error estimates.  

\begin{figure}[ht!]
\includegraphics[width=\columnwidth]{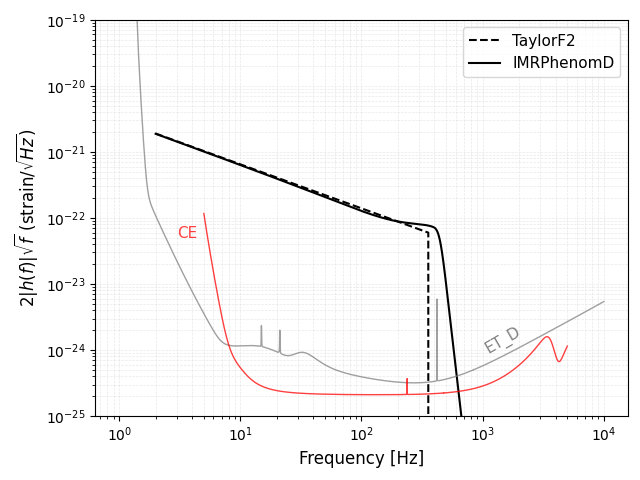}
\caption{Comparing \TAYF and \IMRD waveforms of Signal 2 as observed with one of the components of ET. The signal spectra are plotted in units of Hz$^{-1/2}$ to have them in one plot with the sensitivities of Einstein Telescope and Cosmic Explorer. A cut at $4f_{\rm{isco}}$ is applied to the \TAYF waveform.}
\label{fig:wf_cfr}
\end{figure}
To highlight the model differences we display in Figure \ref{fig:wf_cfr} the normalized amplitudes for the two waveforms for Signal 2 (see Table \ref{tab:signal_sum}) adjusted to units Hz$^{-1/2}$ in comparison also with the ET sensitivity curve. It can be observed that the \IMRD model shows the characteristic bump towards the merger frequency and provides a more natural cut off at high frequency.

\begin{figure}[ht!]
\includegraphics[width=\columnwidth]{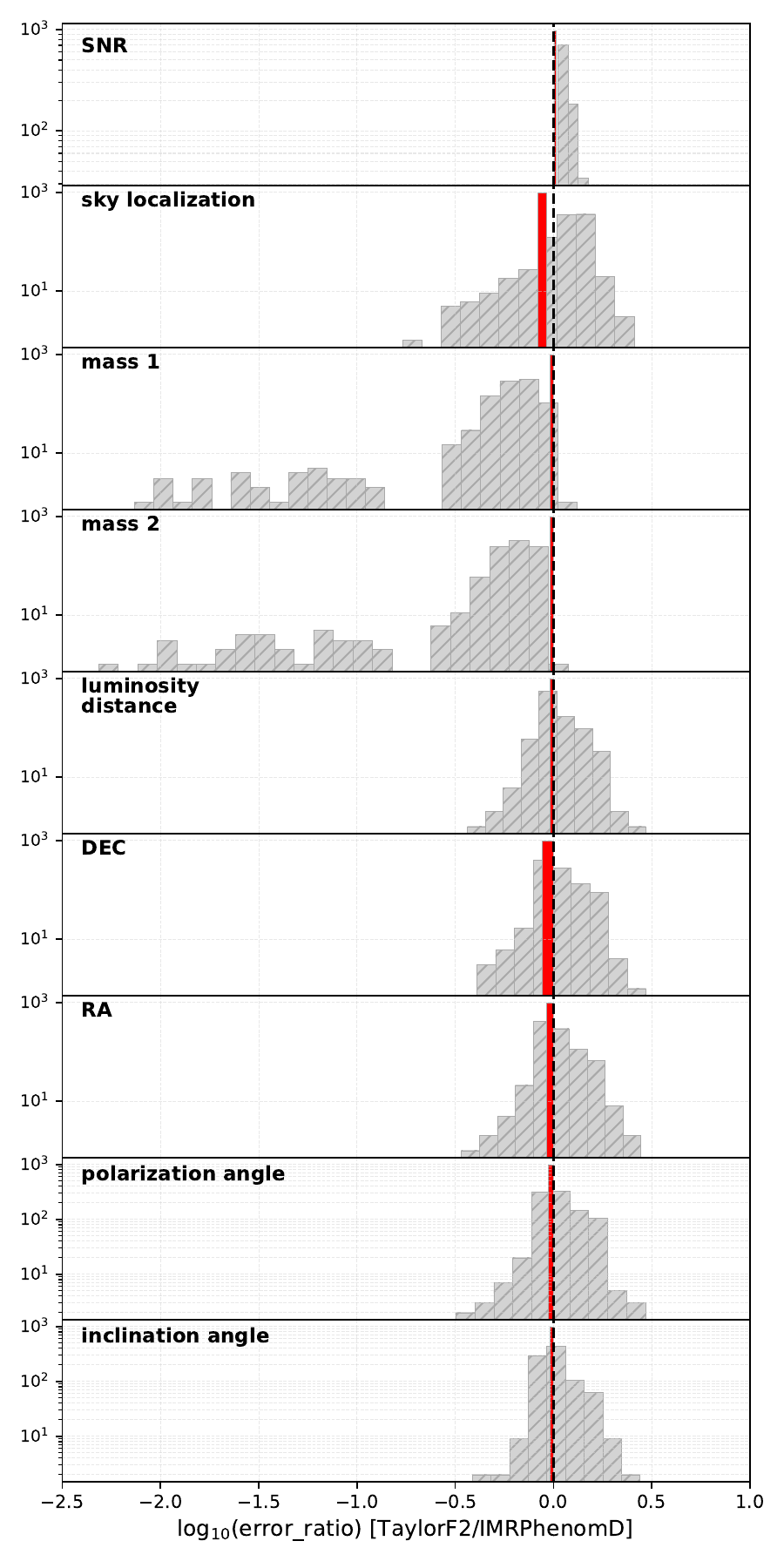}
\caption{Comparison between \TAYF and \IMRD approximants for BBH signals (in shaded gray) and for BNS signals (in solid red). We analyze 927 BBH signals and 966 BNS systems observed by ET with SNR $>$ 8. The frequency range is from 2\,Hz to 1024\,Hz. The frequency resolution used is 1/32\,Hz. We applied a cut at $4f_{\rm{isco}}$ for the \TAYF waveform. The upper histogram shows the ratio between the SNRs. The histograms below show the ratio of the PE errors. The vertical dashed black line marks the equal ratio case.}
\label{fig:wf_err_cfr}
\end{figure}

To compare the two approximants, we start with 1000 BBH signals and 1000 BNS signals. We constrain the BNS redshifts to a maximum of 0.4 so that the SNR distributions for BBHs and BNSs are similar. From these, 927 BBH and 966 BNS signals are seen with a minimum SNR of 8.0 using \TAYF, \IMRD, or \IMRX LALSimulation approximants. The distribution of SNR values of detected signals goes up to 200 peaking at 20. The signals are analyzed with a frequency resolution of 1/32\,Hz. The analysis is carried out with ET alone over a frequency range from 2\,Hz to 1024\,Hz. The results are shown in Figure \ref{fig:wf_err_cfr}. We plot the ratios of the PE errors recovered for each of the parameters with \TAYF and \IMRD as histograms over the detected signals. The BBH results are in gray, while the BNS ones are plotted in red. The upper graph shows the ratio of SNRs, while the other histograms show the ratio of the uncertainties on the other parameters. For BBH systems, the PE errors of the individual masses are underestimated with \TAYF (assuming that \IMRD generally yields more accurate results), while all other parameters considered here are weakly biased towards larger errors compared to \IMRD. For BNS signals there is no significant difference between the two considered approximants: the SNRs are coincident, and only the sky-localisation errors are slightly underestimated with \TAYF. 

\begin{figure}[ht!]
\includegraphics[width=\columnwidth]{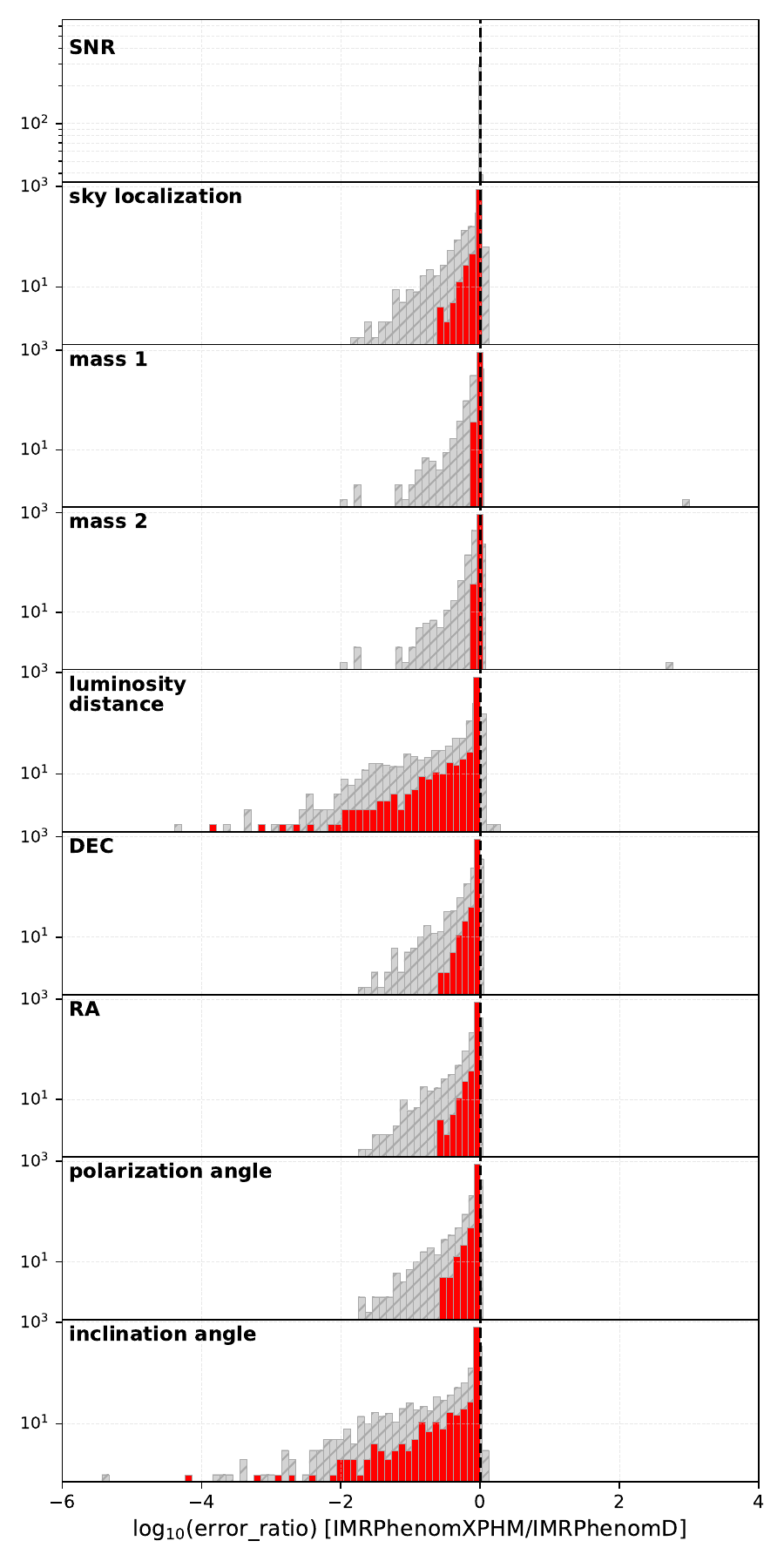}
\caption{Comparison between \IMRX and \IMRD approximants for BBH signals (in shaded gray) and for BNS signals (in solid red). Settings and number of signals are the same as in Figure \ref{fig:wf_err_cfr}.}
\label{fig:wf_err_hom}
\end{figure}

We show a similar analysis for the \IMRX and \IMRD approximants. The \IMRX approximant can describe precession (not considered in this study) and higher-order modes. The latter is well-known to have important consequences for BBH analyses (see, e.g., \cite{BrSe2007,KhEA2020}), and break the distance inclination degeneracy \cite{UMF2019,CVN2019}. Here, a significant improvement of PE can also be seen for a significant fraction of BNS signals. For about 5\% of BNS, the sky-localization error is significantly reduced by up to a factor 5 using \IMRX. For about 10\% of BNS, the distance error is significantly reduced by up to a few orders of magnitude. What this means is that some BNS signals show strong correlations in the likelihood between distance and inclination angle using \IMRD, and the higher-order modes described with \IMRX lead to a suppression of these correlations.

\subsection{Reference studies}
\label{sec:examples}
In the following, we present a few reference PE results obtained from \textsc{gwfish}, which can be used for comparison with other simulation software. We focus on sky-localization results, since these are more difficult to simulate accurately. A single detector can accurately localize a GW source if the observation is long enough so that the motion of the detector leads to amplitude and phase modulations of GW signals, which depend on the direction of the source. This is the case for BNSs, which are observed for several hours with ET. Even lighter NSBHs and BBHs can be localized this way by ET. For specific signals as well as for the average localization by Earth rotation, the localization errors depend on the latitude of the GW detector. Sky-localization is greatly improved if additional detectors are present to form a network. In the latter case, the information comes from measurements of phase differences between signals observed in different detectors (standard triangulation technique) \cite{WeCh2010}. 

\begin{figure*}[ht!]
\includegraphics[width=0.31\textwidth]{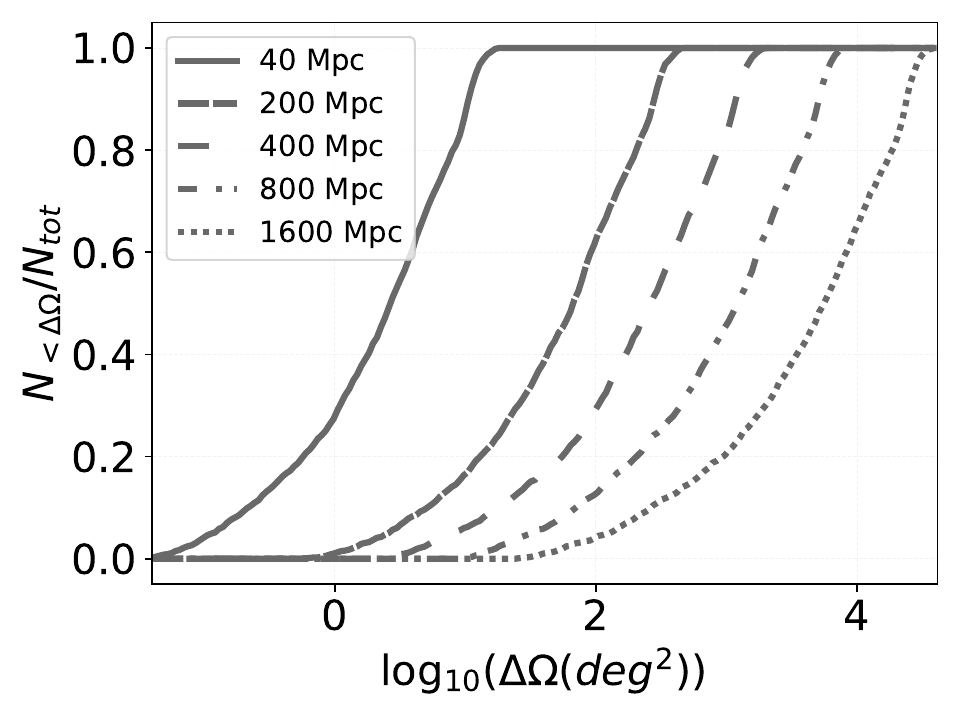}
\includegraphics[width=0.31\textwidth]{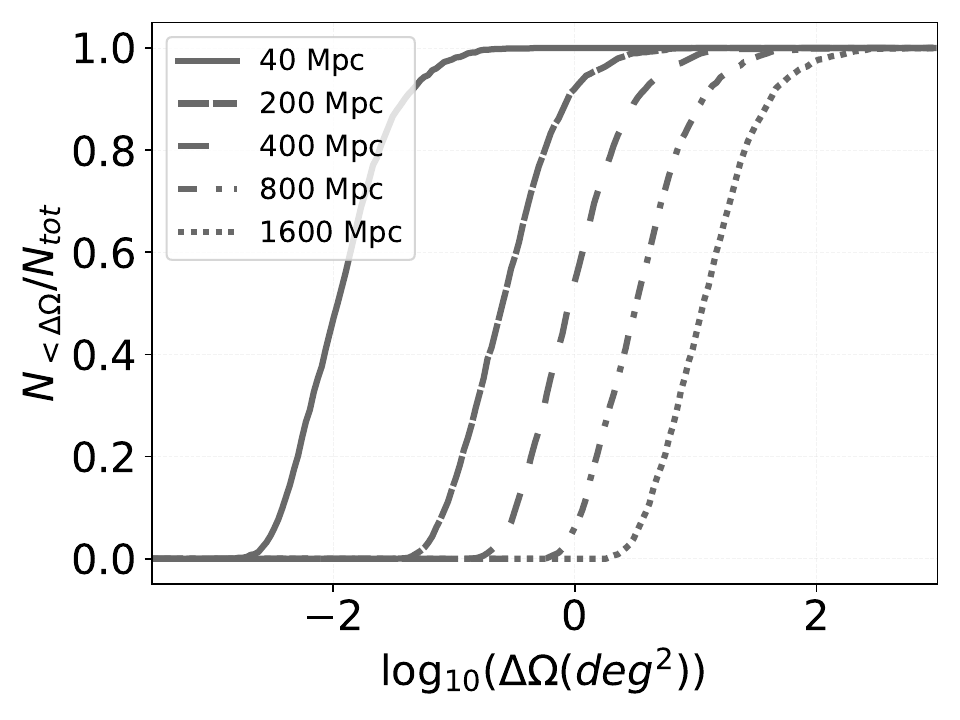}
\includegraphics[width=0.31\textwidth]{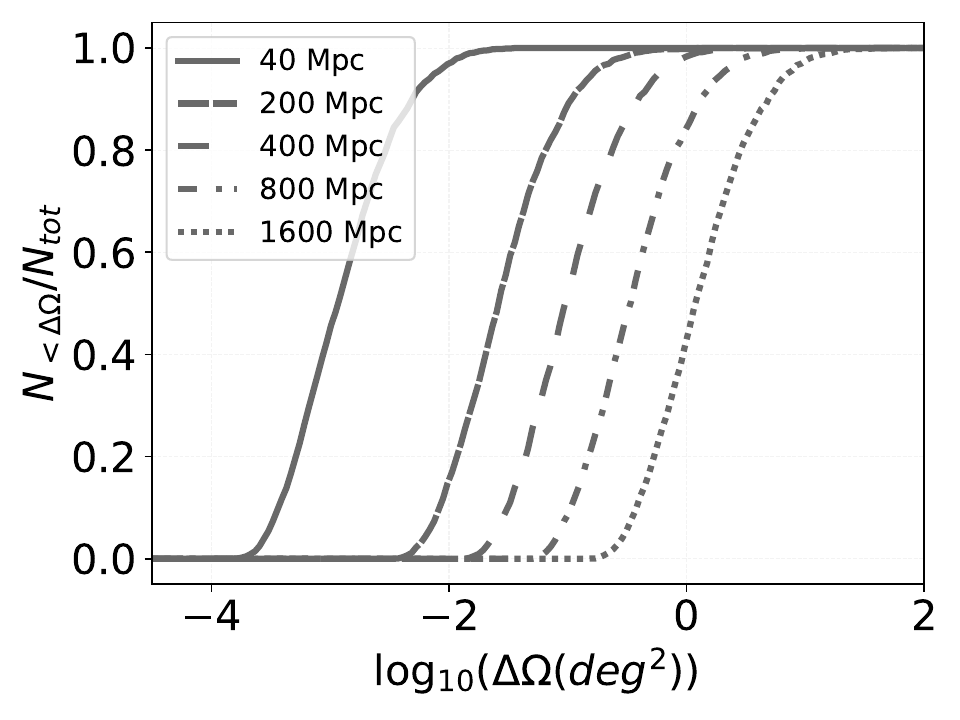}
\caption{Sky-localization capabilities (90\% credible regions) of future detector networks including ET and CE. We used the LIGO sites as locations for the two CE detectors. Left: ET alone, middle: ET and one CE detector at the LIGO Hanford site, right: ET with two CE detectors at the two LIGO sites. The sensitivity models used here is the ET sensitivity formerly known as ET-D \cite{HiEA2011}, and the 40\,km CE sensitivity \cite{EvEA2021}.}
\label{fig:bns_loc}
\end{figure*}

We show in Figure \ref{fig:bns_loc} the sky-localization results for BNSs as $90\%$ credible region. The results are presented as fraction of signals localized within a certain area in the sky. BNSs are considered at various fixed luminosity distances; 1000 BNS per curve. The BNS masses are set to 1.4+1.4$M_\odot$. The plots confirm that ET has unique sky-localization capabilities for BNSs, which would enable multi-messenger studies without the need of other detectors, but it is also clear that the localization capabilities of a network are far superior even if only one other detector like Cosmic Explorer (CE) is added. Nearer systems are better localized and the network of ET+2CE manages to localize almost all of the injected signals within 10\,deg$^2$. We find our results to be consistent with previously published results, e.g., \cite{ChEA2018,Li_2022}.

Next, we present an analysis of the 85 signals of GWTC-3 \cite{AbEA2019c,AbEA2021,AbEA2021a}. Only the merger time, masses and distance were taken from the catalogue. Angular parameters were drawn randomly, spins were set to zero. Observation of these signals with ET was simulated. The simulation was done for 10 realizations of the angular waveform parameters for each signal. Table \ref{tab:GWTC3} provides a summary of the median SNRs and sky-localization that would have been achieved with ET. The table only contains the signals whose sky-localization was within 1000\,deg$^2$ for all random realizations of the angular variables of the waveform. As expected, these signals are among the lowest-mass systems in GWTC-3, which means that ET would have observed them for a long time. Also a low-mass BBH system like GW190707 could potentially be detected before the merger occurs.
\begin{table}[ht!]
\begin{tabular}{|p{3cm}|l|l|l|}
\hline
& GW170817 & GW190425 & GW190707 \\
\hline
Median SNR & 710 & 190 & 230 \\
\hline
Median sky-location error [deg$^2$] (90\% conf.) & 4.0 & 67 & 34 \\
\hline
Masses [$M_\odot$] & 1.46, 1.27 & 2, 1.4 & 12, 8.4 \\
\hline
Luminosity distance [Mpc] & 40 & 160 & 800 \\
\hline
Time until merger starting from 3\,Hz [s] & $2.6\cdot 10^4$ & $1.8\cdot 10^4$ & 930 \\
\hline
\end{tabular}
\caption{Median SNRs and sky-localization errors for simulated observations of GWTC-3 signals with ET. Only masses, merger times, and distances were taken from GWTC-3. The sky-localization errors were evaluated for several random realizations, and only the signals with error $<1000$\,deg$^2$ for all realizations are shown in the table. The waveform used here is \IMRX to include the potentially important higher-order modes, but for the signals reported in the table, the sky-localization results would have been the same using \TAYF or \IMRD. The localization even of larger BBHs can be  $<100$\,deg$^2$, but it depends strongly on the values of angular parameters.}
\label{tab:GWTC3}
\end{table}

\subsection{Multi-band study: ET + LISA}
Here, we present a multi-band study of ET together with LISA, whose observation band extends from about 0.1\,mHz to 0.1\,Hz. We study an IMBH system of $4000M_{\odot}$ and $2000M_{\odot}$ in the source frame at a distance of 1.2\,Gpc, which can be observed by both detectors. The masses are chosen large enough such that the signal traverses the full LISA band within a few years. Other parameter values are the same as for Signal 1. However, we added the remaining four spin parameters (\texttt{tilt$\_$1}$=0.91$, \texttt{tilt$\_$2}$=1.8$, \texttt{phi$\_$12}$=6.1$ and \texttt{phi$\_$jl}$=4.6$) and used the \texttt{IMRPhenomXPHM} waveform approximant, since we anticipated that higher-order modes and precession might play a role. All the 15 parameters were analyzed. Table \ref{tab:multi_band} reports the results of the analysis with \textsc{gwfish}. We can see that LISA provides better information about the component masses, the source distance and the sky localization, while the errors on the angular parameters are similar. Interestingly, when the two detectors are combined in a multi-band observation, then the sky-localization error and the errors on masses and luminosity distance are rather reduced compared to the individual errors thanks to the dominating role of LISA. The opposite happens to the time parameter, which is much better estimated with ET alone. Most of the other errors are reduced more strongly (almost one order of magnitude) than what one would expect from averaging independent errors.
\begin{table}
\centering
\caption{Multi-band observation with ET+LISA. We simulated an IMBH binary system of $4000M_{\odot}$ and $2000M_{\odot}$ at a distance of $1.2$\,Gpc. The errors are given at $1\sigma$ and the sky localization is at $90\%$ C.L.. We used a frequency resolution of 1/16\,Hz for ET and $10^{-4}$\,Hz for LISA.}
\scriptsize{
\begin{tabular}{r|c|c|c|c}
\toprule
&{\textbf{inj\_value}}&\textbf{err\_LISA} &\textbf{err\_ET} &\textbf{err\_LISA\_ET} \\
\hline
\hline
SNR &   & $70.5$ & $250$ & $260$\\
\hline
$m_{1,\rm src}$ [$M_{\odot}$] & $4000$ & $\pm 0.036$ & $\pm 19$ & $\pm 0.032$\\
\hline
$m_{2,\rm src}$ [$M_{\odot}$] & $2000$ & $\pm 0.016$ & $\pm 21$ & $\pm 0.015$\\
\hline
$d_{\rm L}$ [Mpc] & 1200 & $\pm 21$ & $\pm 218$ & $\pm 19$\\
\hline
RA [rad] & 0.37 & $\pm 0.022$ & $\pm 0.029$ & $\pm 0.0072$\\
\hline
DEC [rad] & -0.36 & $\pm 0.0082$ & $\pm 0.046$ & $\pm 0.0054$\\
\hline
$\iota$ [rad] & $1.6$ & $\pm 0.010$ & $\pm 0.040$ & $\pm 0.0081$\\
\hline
$\Psi$ [rad] & 1.7 & $\pm 0.015$ & $\pm 0.040$ & $\pm 0.0054$\\
\hline
phase [rad] & 3.5 & $\pm 0.72$ & $\pm 0.11$ & $\pm 0.028$\\
\hline
$t_c$ [s] & 1120381489.6 & $\pm 0.59$ & $\pm 0.0050$ & $\pm 0.0020$\\
\hline
sky-loc [deg$^2$] & & 7.2 & 56 & 1.1\\
\hline 
\end{tabular}
}
\label{tab:multi_band}
\end{table}

\section{Conclusions}
In this paper, we have presented \textsc{gwfish}; a new Fisher-matrix code to investigate parameter-estimation capabilities of future GW detectors. \textsc{gwfish} stands out as a robust and flexible detector simulation software: 
\begin{itemize}
    \item It implements modeling of the so-called $A, E, T$ channels of LISA-type space-borne detectors, of LGWA, and of triangular and L-shape terrestrial GW detectors;
    \item It can use all LALSimulation approximants supporting precession as well as tidal deformation;
    \item It incorporates the motion of detectors, e.g., because of Earth/Moon rotation, or the orbits of satellites around the Sun, and it simulates the corresponding amplitude and phase modulations of GW signals;
    \item It can simulate multi-band observations;
    \item It takes several measures to reduce the impact of numerical errors on Fisher-matrix PE errors.
\end{itemize}
Certain analyses are currently not possible with state-of-the-art detector simulation and Bayesian analysis software like \textsc{bilby} due to practical constraints on available computational resources. For example, if hundreds of thousands of GW signals need to be analyzed, sometimes repeatedly to study the effect of changes in population or waveform models, then Fisher-matrix codes like \textsc{gwfish} are the best option. Important is to understand the limitations of Gaussian approximations of the likelihood. We have shown a few comparisons in this article between \textsc{bilby} and \textsc{gwfish}, and we generally find a good match between the two with exceptions that can be explained. In the cases studied here, discrepancies emerged when multiple modes of the posterior distribution contributed significantly to the PE errors. Tighter priors would significantly reduce the mismatch, as we have seen in the case of Signal 1.

We also investigated the impact of the choice of waveform model, i.e., \TAYF vs \IMRD vs \IMRX. For BBH signals, all three approximants produce significantly different results. For the BNS analyses presented here, namely SNR and sky-localization, \TAYF vs \IMRD produce almost identical results. However, for a significant fraction of BNS signals, PE errors with \IMRX are significantly reduced compared to \IMRD. The reduction is strongest for distance errors (10\% of signals significantly affected), because distance errors can be strongly correlated with errors of the inclination angle. The \IMRX approximant takes significantly longer to evaluate than \TAYF and \IMRD (the latter two with similar evaluation times) when using the LALSimulation package with Python/SWIG \cite{Wet2020}. We conclude that the choice of approximant is generally more important for BBH analyses. For BNS signals, broad population studies can still be carried out with \IMRD or \TAYF, but results especially for distance and inclination errors are more reliable for a significant fraction of signals when using \IMRX. Needless to say, it always depends on the type of analysis one wants to perform whether a certain waveform model is adequate or not. Precession and higher-order modes might have negligible effects in some analyses, and be crucial in others. The \textsc{gwfish} user must be familiar with the characteristics of the standard waveform models.

The scientific results presented in this paper confirm the long-standing claim that the triangular configuration of ET makes it possible to measure GW polarizations and thereby break certain model degeneracies. For example, for high-SNR BBH signals, we found that the inclination angle is measured precisely with ET, which also improves the estimates of source distance. Instead, when the SNR is rather low, as in the SNR=8 case, the Fisher matrix analysis starts to break down. We also found that ET itself would have estimated the sky location of the low-mass GWTC-3 signals to within 100\,deg$^2$ at 90\% confidence (depending on the exact sky location, the localization of GW170817 can be within 1\,deg$^2$). This shows that multi-messenger astronomy with ET as sole GW detector is possible, but unlocking the full potential of multi-messenger astronomy with GWs would require at least another GW detector with a similar sensitivity.

The next developments of \textsc{gwfish} include the implementation of posterior-sampling algorithms starting from Gaussian likelihoods with priors. This will ameliorate one of the strongest short-comings of Fisher-matrix analyses, and at the same time still avoid the large computational cost of a full Bayesian analysis. It is also planned to include a greater variety of GW sources in the simulation, e.g., white-dwarf binaries, intermediate-mass ratio inspirals, tidal-disruption events, etc, which will make \textsc{gwfish} more useful for simulations of space-based and lunar GW detectors.

\section*{Acknowledgments}
We thank Michele Mancarella, Francesco Iacovelli, Stefano Foffa, and Michele Maggiore for discussions about the Fisher-matrix technique and for comparisons of results with their simulation code. We thank Stefano Bagnasco and the IT team of University of Turin for support with computational resources. BG is supported by the Italian Ministry of Education, University and Research within the PRIN 2017 Research Program Framework, n. 2017SYRTCN. MB, AM and SR acknowledge financial support from MIUR (PRIN 2020 grant 2020KB33TP\_001). BB acknowledges financial support from MIUR (PRIN 2017 grant 20179ZF5KS).

\raggedright
\bibliographystyle{apsrev}
\bibliography{references}

\begin{thebibliography}{57}
\expandafter\ifx\csname natexlab\endcsname\relax\def\natexlab#1{#1}\fi
\expandafter\ifx\csname bibnamefont\endcsname\relax
  \def\bibnamefont#1{#1}\fi
\expandafter\ifx\csname bibfnamefont\endcsname\relax
  \def\bibfnamefont#1{#1}\fi
\expandafter\ifx\csname citenamefont\endcsname\relax
  \def\citenamefont#1{#1}\fi
\expandafter\ifx\csname url\endcsname\relax
  \def\url#1{\texttt{#1}}\fi
\expandafter\ifx\csname urlprefix\endcsname\relax\def\urlprefix{URL }\fi
\providecommand{\bibinfo}[2]{#2}
\providecommand{\eprint}[2][]{\url{#2}}

\bibitem[{\citenamefont{Maggiore et~al.}(2020)\citenamefont{Maggiore, Broeck, Bartolo, Belgacem, Bertacca, Bizouard, Branchesi, Clesse, Foffa, Garc{\'{\i}}a-Bellido et~al.}}]{MaEA2020}
\bibinfo{author}{\bibfnamefont{M.}~\bibnamefont{Maggiore}}, \bibinfo{author}{\bibfnamefont{C.~V.~D.} \bibnamefont{Broeck}}, \bibinfo{author}{\bibfnamefont{N.}~\bibnamefont{Bartolo}}, \bibinfo{author}{\bibfnamefont{E.}~\bibnamefont{Belgacem}}, \bibinfo{author}{\bibfnamefont{D.}~\bibnamefont{Bertacca}}, \bibinfo{author}{\bibfnamefont{M.~A.} \bibnamefont{Bizouard}}, \bibinfo{author}{\bibfnamefont{M.}~\bibnamefont{Branchesi}}, \bibinfo{author}{\bibfnamefont{S.}~\bibnamefont{Clesse}}, \bibinfo{author}{\bibfnamefont{S.}~\bibnamefont{Foffa}}, \bibinfo{author}{\bibfnamefont{J.}~\bibnamefont{Garc{\'{\i}}a-Bellido}}, \bibnamefont{et~al.}, \bibinfo{journal}{Journal of Cosmology and Astroparticle Physics} \textbf{\bibinfo{volume}{2020}}, \bibinfo{pages}{050} (\bibinfo{year}{2020}), \urlprefix\url{https://doi.org/10.1088\%2F1475-7516\%2F2020\%2F03\%2F050}.

\bibitem[{\citenamefont{Kawamura et~al.}(2021)\citenamefont{Kawamura, Ando, Seto, Sato, Musha, Kawano, Yokoyama, Tanaka, Ioka, Akutsu et~al.}}]{KaEA2021}
\bibinfo{author}{\bibfnamefont{S.}~\bibnamefont{Kawamura}}, \bibinfo{author}{\bibfnamefont{M.}~\bibnamefont{Ando}}, \bibinfo{author}{\bibfnamefont{N.}~\bibnamefont{Seto}}, \bibinfo{author}{\bibfnamefont{S.}~\bibnamefont{Sato}}, \bibinfo{author}{\bibfnamefont{M.}~\bibnamefont{Musha}}, \bibinfo{author}{\bibfnamefont{I.}~\bibnamefont{Kawano}}, \bibinfo{author}{\bibfnamefont{J.}~\bibnamefont{Yokoyama}}, \bibinfo{author}{\bibfnamefont{T.}~\bibnamefont{Tanaka}}, \bibinfo{author}{\bibfnamefont{K.}~\bibnamefont{Ioka}}, \bibinfo{author}{\bibfnamefont{T.}~\bibnamefont{Akutsu}}, \bibnamefont{et~al.}, \bibinfo{journal}{Progress of Theoretical and Experimental Physics} \textbf{\bibinfo{volume}{2021}} (\bibinfo{year}{2021}), ISSN \bibinfo{issn}{2050-3911}, \urlprefix\url{https://doi.org/10.1093/ptep/ptab019}.

\bibitem[{\citenamefont{Acernese et~al.}(2014)\citenamefont{Acernese, Agathos, Agatsuma, Aisa, Allemandou, Allocca, Amarni, Astone, Balestri, Ballardin et~al.}}]{AcEA2015}
\bibinfo{author}{\bibfnamefont{F.}~\bibnamefont{Acernese}}, \bibinfo{author}{\bibfnamefont{M.}~\bibnamefont{Agathos}}, \bibinfo{author}{\bibfnamefont{K.}~\bibnamefont{Agatsuma}}, \bibinfo{author}{\bibfnamefont{D.}~\bibnamefont{Aisa}}, \bibinfo{author}{\bibfnamefont{N.}~\bibnamefont{Allemandou}}, \bibinfo{author}{\bibfnamefont{A.}~\bibnamefont{Allocca}}, \bibinfo{author}{\bibfnamefont{J.}~\bibnamefont{Amarni}}, \bibinfo{author}{\bibfnamefont{P.}~\bibnamefont{Astone}}, \bibinfo{author}{\bibfnamefont{G.}~\bibnamefont{Balestri}}, \bibinfo{author}{\bibfnamefont{G.}~\bibnamefont{Ballardin}}, \bibnamefont{et~al.}, \bibinfo{journal}{Classical and Quantum Gravity} \textbf{\bibinfo{volume}{32}}, \bibinfo{pages}{024001} (\bibinfo{year}{2014}), \urlprefix\url{https://doi.org/10.1088%2F0264-9381%2F32%2F2%2F024001}.

\bibitem[{\citenamefont{Aasi et~al.}(2015)\citenamefont{Aasi, Abbott, Abbott, Abbott, Abernathy, Ackley, Adams, Adams, Addesso, Adhikari et~al.}}]{LSC2015}
\bibinfo{author}{\bibfnamefont{J.}~\bibnamefont{Aasi}}, \bibinfo{author}{\bibfnamefont{B.~P.} \bibnamefont{Abbott}}, \bibinfo{author}{\bibfnamefont{R.}~\bibnamefont{Abbott}}, \bibinfo{author}{\bibfnamefont{T.}~\bibnamefont{Abbott}}, \bibinfo{author}{\bibfnamefont{M.~R.} \bibnamefont{Abernathy}}, \bibinfo{author}{\bibfnamefont{K.}~\bibnamefont{Ackley}}, \bibinfo{author}{\bibfnamefont{C.}~\bibnamefont{Adams}}, \bibinfo{author}{\bibfnamefont{T.}~\bibnamefont{Adams}}, \bibinfo{author}{\bibfnamefont{P.}~\bibnamefont{Addesso}}, \bibinfo{author}{\bibfnamefont{R.~X.} \bibnamefont{Adhikari}}, \bibnamefont{et~al.}, \bibinfo{journal}{Classical and Quantum Gravity} \textbf{\bibinfo{volume}{32}}, \bibinfo{pages}{074001} (\bibinfo{year}{2015}), \urlprefix\url{https://doi.org/10.1088\%2F0264-9381\%2F32\%2F7\%2F074001}.

\bibitem[{\citenamefont{Souradeep}(2016)}]{Sou2016}
\bibinfo{author}{\bibfnamefont{T.}~\bibnamefont{Souradeep}}, \bibinfo{journal}{Resonance} \textbf{\bibinfo{volume}{21}}, \bibinfo{pages}{225 } (\bibinfo{year}{2016}), \urlprefix\url{https://doi.org/10.1007/s12045-016-0316-6}.

\bibitem[{\citenamefont{Akutsu et~al.}(2019)\citenamefont{Akutsu, Ando, Arai, Arai, Araki, Araya, Aritomi, Asada, Aso, Atsuta et~al.}}]{AkEA2018}
\bibinfo{author}{\bibfnamefont{T.}~\bibnamefont{Akutsu}}, \bibinfo{author}{\bibfnamefont{M.}~\bibnamefont{Ando}}, \bibinfo{author}{\bibfnamefont{K.}~\bibnamefont{Arai}}, \bibinfo{author}{\bibfnamefont{Y.}~\bibnamefont{Arai}}, \bibinfo{author}{\bibfnamefont{S.}~\bibnamefont{Araki}}, \bibinfo{author}{\bibfnamefont{A.}~\bibnamefont{Araya}}, \bibinfo{author}{\bibfnamefont{N.}~\bibnamefont{Aritomi}}, \bibinfo{author}{\bibfnamefont{H.}~\bibnamefont{Asada}}, \bibinfo{author}{\bibfnamefont{Y.}~\bibnamefont{Aso}}, \bibinfo{author}{\bibfnamefont{S.}~\bibnamefont{Atsuta}}, \bibnamefont{et~al.}, \bibinfo{journal}{Nature Astronomy} \textbf{\bibinfo{volume}{3}}, \bibinfo{pages}{35} (\bibinfo{year}{2019}), ISSN \bibinfo{issn}{2397-3366}, \urlprefix\url{https://www.nature.com/articles/s41550-018-0658-y}.

\bibitem[{\citenamefont{Punturo et~al.}(2010)\citenamefont{Punturo, Abernathy, Acernese, Allen, Andersson, Arun, Barone, Barr, Barsuglia, Beker et~al.}}]{PuEA2010}
\bibinfo{author}{\bibfnamefont{M.}~\bibnamefont{Punturo}}, \bibinfo{author}{\bibfnamefont{M.}~\bibnamefont{Abernathy}}, \bibinfo{author}{\bibfnamefont{F.}~\bibnamefont{Acernese}}, \bibinfo{author}{\bibfnamefont{B.}~\bibnamefont{Allen}}, \bibinfo{author}{\bibfnamefont{N.}~\bibnamefont{Andersson}}, \bibinfo{author}{\bibfnamefont{K.}~\bibnamefont{Arun}}, \bibinfo{author}{\bibfnamefont{F.}~\bibnamefont{Barone}}, \bibinfo{author}{\bibfnamefont{B.}~\bibnamefont{Barr}}, \bibinfo{author}{\bibfnamefont{M.}~\bibnamefont{Barsuglia}}, \bibinfo{author}{\bibfnamefont{M.}~\bibnamefont{Beker}}, \bibnamefont{et~al.}, \bibinfo{journal}{Classical and Quantum Gravity} \textbf{\bibinfo{volume}{27}}, \bibinfo{pages}{194002} (\bibinfo{year}{2010}), \urlprefix\url{http://stacks.iop.org/0264-9381/27/i=19/a=194002}.

\bibitem[{\citenamefont{{ET Steering Committee}}(2020)}]{ET2020}
\bibinfo{author}{\bibnamefont{{ET Steering Committee}}}, \bibinfo{journal}{{available from European Gravitational Observatory, document number ET-0007B-20}}  (\bibinfo{year}{2020}), \urlprefix\url{https://apps.et-gw.eu/tds/ql/?c=15418}.

\bibitem[{\citenamefont{Evans et~al.}(2021)\citenamefont{Evans, Adhikari, Afle, Ballmer, Biscoveanu, Borhanian, Brown, Chen, Eisenstein, Gruson et~al.}}]{EvEA2021}
\bibinfo{author}{\bibfnamefont{M.}~\bibnamefont{Evans}}, \bibinfo{author}{\bibfnamefont{R.~X.} \bibnamefont{Adhikari}}, \bibinfo{author}{\bibfnamefont{C.}~\bibnamefont{Afle}}, \bibinfo{author}{\bibfnamefont{S.~W.} \bibnamefont{Ballmer}}, \bibinfo{author}{\bibfnamefont{S.}~\bibnamefont{Biscoveanu}}, \bibinfo{author}{\bibfnamefont{S.}~\bibnamefont{Borhanian}}, \bibinfo{author}{\bibfnamefont{D.~A.} \bibnamefont{Brown}}, \bibinfo{author}{\bibfnamefont{Y.}~\bibnamefont{Chen}}, \bibinfo{author}{\bibfnamefont{R.}~\bibnamefont{Eisenstein}}, \bibinfo{author}{\bibfnamefont{A.}~\bibnamefont{Gruson}}, \bibnamefont{et~al.}, \emph{\bibinfo{title}{A horizon study for cosmic explorer: Science, observatories, and community}} (\bibinfo{year}{2021}), \urlprefix\url{https://arxiv.org/abs/2109.09882}.

\bibitem[{\citenamefont{{Amaro-Seoane} et~al.}(2017)\citenamefont{{Amaro-Seoane}, {Audley}, {Babak}, {Baker}, {Barausse}, {Bender}, {Berti}, {Binetruy}, {Born}, {Bortoluzzi} et~al.}}]{ASEA2017}
\bibinfo{author}{\bibfnamefont{P.}~\bibnamefont{{Amaro-Seoane}}}, \bibinfo{author}{\bibfnamefont{H.}~\bibnamefont{{Audley}}}, \bibinfo{author}{\bibfnamefont{S.}~\bibnamefont{{Babak}}}, \bibinfo{author}{\bibfnamefont{J.}~\bibnamefont{{Baker}}}, \bibinfo{author}{\bibfnamefont{E.}~\bibnamefont{{Barausse}}}, \bibinfo{author}{\bibfnamefont{P.}~\bibnamefont{{Bender}}}, \bibinfo{author}{\bibfnamefont{E.}~\bibnamefont{{Berti}}}, \bibinfo{author}{\bibfnamefont{P.}~\bibnamefont{{Binetruy}}}, \bibinfo{author}{\bibfnamefont{M.}~\bibnamefont{{Born}}}, \bibinfo{author}{\bibfnamefont{D.}~\bibnamefont{{Bortoluzzi}}}, \bibnamefont{et~al.}, \bibinfo{journal}{arXiv e-prints}  (\bibinfo{year}{2017}), \urlprefix\url{https://arxiv.org/abs/1702.00786}.

\bibitem[{\citenamefont{Hall et~al.}(2021)\citenamefont{Hall, Kuns, Smith, Bai, Wipf, Biscans, Adhikari, Arai, Ballmer, Barsotti et~al.}}]{HaEA2021}
\bibinfo{author}{\bibfnamefont{E.~D.} \bibnamefont{Hall}}, \bibinfo{author}{\bibfnamefont{K.}~\bibnamefont{Kuns}}, \bibinfo{author}{\bibfnamefont{J.~R.} \bibnamefont{Smith}}, \bibinfo{author}{\bibfnamefont{Y.}~\bibnamefont{Bai}}, \bibinfo{author}{\bibfnamefont{C.}~\bibnamefont{Wipf}}, \bibinfo{author}{\bibfnamefont{S.}~\bibnamefont{Biscans}}, \bibinfo{author}{\bibfnamefont{R.~X.} \bibnamefont{Adhikari}}, \bibinfo{author}{\bibfnamefont{K.}~\bibnamefont{Arai}}, \bibinfo{author}{\bibfnamefont{S.}~\bibnamefont{Ballmer}}, \bibinfo{author}{\bibfnamefont{L.}~\bibnamefont{Barsotti}}, \bibnamefont{et~al.}, \bibinfo{journal}{Phys. Rev. D} \textbf{\bibinfo{volume}{103}}, \bibinfo{pages}{122004} (\bibinfo{year}{2021}), \urlprefix\url{https://link.aps.org/doi/10.1103/PhysRevD.103.122004}.

\bibitem[{\citenamefont{Sesana}(2016)}]{Ses2016}
\bibinfo{author}{\bibfnamefont{A.}~\bibnamefont{Sesana}}, \bibinfo{journal}{Phys. Rev. Lett.} \textbf{\bibinfo{volume}{116}}, \bibinfo{pages}{231102} (\bibinfo{year}{2016}), \urlprefix\url{https://link.aps.org/doi/10.1103/PhysRevLett.116.231102}.

\bibitem[{\citenamefont{Vitale}(2016)}]{Vit2016}
\bibinfo{author}{\bibfnamefont{S.}~\bibnamefont{Vitale}}, \bibinfo{journal}{Phys. Rev. Lett.} \textbf{\bibinfo{volume}{117}}, \bibinfo{pages}{051102} (\bibinfo{year}{2016}), \urlprefix\url{https://link.aps.org/doi/10.1103/PhysRevLett.117.051102}.

\bibitem[{\citenamefont{Grimm and Harms}(2020)}]{GrHa2020}
\bibinfo{author}{\bibfnamefont{S.}~\bibnamefont{Grimm}} \bibnamefont{and} \bibinfo{author}{\bibfnamefont{J.}~\bibnamefont{Harms}}, \bibinfo{journal}{Phys. Rev. D} \textbf{\bibinfo{volume}{102}}, \bibinfo{pages}{022007} (\bibinfo{year}{2020}), \urlprefix\url{https://link.aps.org/doi/10.1103/PhysRevD.102.022007}.

\bibitem[{\citenamefont{Chan et~al.}(2018)\citenamefont{Chan, Messenger, Heng, and Hendry}}]{ChEA2018}
\bibinfo{author}{\bibfnamefont{M.~L.} \bibnamefont{Chan}}, \bibinfo{author}{\bibfnamefont{C.}~\bibnamefont{Messenger}}, \bibinfo{author}{\bibfnamefont{I.~S.} \bibnamefont{Heng}}, \bibnamefont{and} \bibinfo{author}{\bibfnamefont{M.}~\bibnamefont{Hendry}}, \bibinfo{journal}{Phys. Rev. D} \textbf{\bibinfo{volume}{97}}, \bibinfo{pages}{123014} (\bibinfo{year}{2018}), \urlprefix\url{https://link.aps.org/doi/10.1103/PhysRevD.97.123014}.

\bibitem[{\citenamefont{Borhanian}(2021)}]{Bor2021}
\bibinfo{author}{\bibfnamefont{S.}~\bibnamefont{Borhanian}}, \bibinfo{journal}{Classical and Quantum Gravity} \textbf{\bibinfo{volume}{38}}, \bibinfo{pages}{175014} (\bibinfo{year}{2021}), \urlprefix\url{https://doi.org/10.1088/1361-6382/ac1618}.

\bibitem[{\citenamefont{Cutler}(1998)}]{Cut1998}
\bibinfo{author}{\bibfnamefont{C.}~\bibnamefont{Cutler}}, \bibinfo{journal}{Phys. Rev. D} \textbf{\bibinfo{volume}{57}}, \bibinfo{pages}{7089} (\bibinfo{year}{1998}), \urlprefix\url{https://link.aps.org/doi/10.1103/PhysRevD.57.7089}.

\bibitem[{\citenamefont{Nitz and Canton}(2021)}]{NiCa2021}
\bibinfo{author}{\bibfnamefont{A.~H.} \bibnamefont{Nitz}} \bibnamefont{and} \bibinfo{author}{\bibfnamefont{T.~D.} \bibnamefont{Canton}}, \bibinfo{journal}{The Astrophysical Journal Letters} \textbf{\bibinfo{volume}{917}}, \bibinfo{pages}{L27} (\bibinfo{year}{2021}), \urlprefix\url{https://doi.org/10.3847/2041-8213/ac1a75}.

\bibitem[{\citenamefont{Vallisneri}(2008)}]{Val2008}
\bibinfo{author}{\bibfnamefont{M.}~\bibnamefont{Vallisneri}}, \bibinfo{journal}{Phys. Rev. D} \textbf{\bibinfo{volume}{77}}, \bibinfo{pages}{042001} (\bibinfo{year}{2008}), \urlprefix\url{https://link.aps.org/doi/10.1103/PhysRevD.77.042001}.

\bibitem[{\citenamefont{Rodriguez et~al.}(2013)\citenamefont{Rodriguez, Farr, Farr, and Mandel}}]{RoEA2013}
\bibinfo{author}{\bibfnamefont{C.~L.} \bibnamefont{Rodriguez}}, \bibinfo{author}{\bibfnamefont{B.}~\bibnamefont{Farr}}, \bibinfo{author}{\bibfnamefont{W.~M.} \bibnamefont{Farr}}, \bibnamefont{and} \bibinfo{author}{\bibfnamefont{I.}~\bibnamefont{Mandel}}, \bibinfo{journal}{Phys. Rev. D} \textbf{\bibinfo{volume}{88}}, \bibinfo{pages}{084013} (\bibinfo{year}{2013}), \urlprefix\url{https://link.aps.org/doi/10.1103/PhysRevD.88.084013}.

\bibitem[{\citenamefont{Armstrong et~al.}(1999)\citenamefont{Armstrong, Estabrook, and Tinto}}]{AET1999}
\bibinfo{author}{\bibfnamefont{J.~W.} \bibnamefont{Armstrong}}, \bibinfo{author}{\bibfnamefont{F.~B.} \bibnamefont{Estabrook}}, \bibnamefont{and} \bibinfo{author}{\bibfnamefont{M.}~\bibnamefont{Tinto}}, \bibinfo{journal}{The Astrophysical Journal} \textbf{\bibinfo{volume}{527}}, \bibinfo{pages}{814} (\bibinfo{year}{1999}), \urlprefix\url{https://doi.org/10.1086/308110}.

\bibitem[{\citenamefont{Prince et~al.}(2002)\citenamefont{Prince, Tinto, Larson, and Armstrong}}]{PrEA2002}
\bibinfo{author}{\bibfnamefont{T.~A.} \bibnamefont{Prince}}, \bibinfo{author}{\bibfnamefont{M.}~\bibnamefont{Tinto}}, \bibinfo{author}{\bibfnamefont{S.~L.} \bibnamefont{Larson}}, \bibnamefont{and} \bibinfo{author}{\bibfnamefont{J.~W.} \bibnamefont{Armstrong}}, \bibinfo{journal}{Phys. Rev. D} \textbf{\bibinfo{volume}{66}}, \bibinfo{pages}{122002} (\bibinfo{year}{2002}), \urlprefix\url{https://link.aps.org/doi/10.1103/PhysRevD.66.122002}.

\bibitem[{\citenamefont{Vallisneri}(2005)}]{Val2005a}
\bibinfo{author}{\bibfnamefont{M.}~\bibnamefont{Vallisneri}}, \bibinfo{journal}{Phys. Rev. D} \textbf{\bibinfo{volume}{71}}, \bibinfo{pages}{022001} (\bibinfo{year}{2005}), \urlprefix\url{https://link.aps.org/doi/10.1103/PhysRevD.71.022001}.

\bibitem[{\citenamefont{Hinderer et~al.}(2010)\citenamefont{Hinderer, Lackey, Lang, and Read}}]{HiEA2010}
\bibinfo{author}{\bibfnamefont{T.}~\bibnamefont{Hinderer}}, \bibinfo{author}{\bibfnamefont{B.~D.} \bibnamefont{Lackey}}, \bibinfo{author}{\bibfnamefont{R.~N.} \bibnamefont{Lang}}, \bibnamefont{and} \bibinfo{author}{\bibfnamefont{J.~S.} \bibnamefont{Read}}, \bibinfo{journal}{Phys. Rev. D} \textbf{\bibinfo{volume}{81}}, \bibinfo{pages}{123016} (\bibinfo{year}{2010}), \urlprefix\url{http://link.aps.org/doi/10.1103/PhysRevD.81.123016}.

\bibitem[{\citenamefont{{Ronchini} et~al.}(2022)\citenamefont{{Ronchini}, {Branchesi}, {Oganesyan}, {Banerjee}, {Dupletsa}, {Ghirlanda}, {Harms}, {Mapelli}, and {Santoliquido}}}]{Ronchini2022}
\bibinfo{author}{\bibfnamefont{S.}~\bibnamefont{{Ronchini}}}, \bibinfo{author}{\bibfnamefont{M.}~\bibnamefont{{Branchesi}}}, \bibinfo{author}{\bibfnamefont{G.}~\bibnamefont{{Oganesyan}}}, \bibinfo{author}{\bibfnamefont{B.}~\bibnamefont{{Banerjee}}}, \bibinfo{author}{\bibfnamefont{U.}~\bibnamefont{{Dupletsa}}}, \bibinfo{author}{\bibfnamefont{G.}~\bibnamefont{{Ghirlanda}}}, \bibinfo{author}{\bibfnamefont{J.}~\bibnamefont{{Harms}}}, \bibinfo{author}{\bibfnamefont{M.}~\bibnamefont{{Mapelli}}}, \bibnamefont{and} \bibinfo{author}{\bibfnamefont{F.}~\bibnamefont{{Santoliquido}}}, \bibinfo{journal}{arXiv e-prints} \bibinfo{eid}{arXiv:2204.01746} (\bibinfo{year}{2022}), \eprint{2204.01746}.

\bibitem[{\citenamefont{Goncharov et~al.}(2022)\citenamefont{Goncharov, Nitz, and Harms}}]{goncharov2022null}
\bibinfo{author}{\bibfnamefont{B.}~\bibnamefont{Goncharov}}, \bibinfo{author}{\bibfnamefont{A.~H.} \bibnamefont{Nitz}}, \bibnamefont{and} \bibinfo{author}{\bibfnamefont{J.}~\bibnamefont{Harms}}, \emph{\bibinfo{title}{Utilizing the null stream of einstein telescope}} (\bibinfo{year}{2022}), \urlprefix\url{https://arxiv.org/abs/2204.08533}.

\bibitem[{\citenamefont{Schilling}(1997)}]{Sch1997}
\bibinfo{author}{\bibfnamefont{R.}~\bibnamefont{Schilling}}, \bibinfo{journal}{Classical and Quantum Gravity} \textbf{\bibinfo{volume}{14}}, \bibinfo{pages}{1513} (\bibinfo{year}{1997}), \urlprefix\url{https://doi.org/10.1088/0264-9381/14/6/020}.

\bibitem[{\citenamefont{Abbott et~al.}(2017)\citenamefont{Abbott, Abbott, Abbott, Abernathy, Ackley, Adams, Addesso, Adhikari, Adya, Affeldt et~al.}}]{AbEA2017a}
\bibinfo{author}{\bibfnamefont{B.~P.} \bibnamefont{Abbott}}, \bibinfo{author}{\bibfnamefont{R.}~\bibnamefont{Abbott}}, \bibinfo{author}{\bibfnamefont{T.~D.} \bibnamefont{Abbott}}, \bibinfo{author}{\bibfnamefont{M.~R.} \bibnamefont{Abernathy}}, \bibinfo{author}{\bibfnamefont{K.}~\bibnamefont{Ackley}}, \bibinfo{author}{\bibfnamefont{C.}~\bibnamefont{Adams}}, \bibinfo{author}{\bibfnamefont{P.}~\bibnamefont{Addesso}}, \bibinfo{author}{\bibfnamefont{R.~X.} \bibnamefont{Adhikari}}, \bibinfo{author}{\bibfnamefont{V.~B.} \bibnamefont{Adya}}, \bibinfo{author}{\bibfnamefont{C.}~\bibnamefont{Affeldt}}, \bibnamefont{et~al.}, \bibinfo{journal}{Classical and Quantum Gravity} \textbf{\bibinfo{volume}{34}}, \bibinfo{pages}{044001} (\bibinfo{year}{2017}), \urlprefix\url{http://stacks.iop.org/0264-9381/34/i=4/a=044001}.

\bibitem[{\citenamefont{Essick et~al.}(2017)\citenamefont{Essick, Vitale, and Evans}}]{EVE2017}
\bibinfo{author}{\bibfnamefont{R.}~\bibnamefont{Essick}}, \bibinfo{author}{\bibfnamefont{S.}~\bibnamefont{Vitale}}, \bibnamefont{and} \bibinfo{author}{\bibfnamefont{M.}~\bibnamefont{Evans}}, \bibinfo{journal}{Phys. Rev. D} \textbf{\bibinfo{volume}{96}}, \bibinfo{pages}{084004} (\bibinfo{year}{2017}), \urlprefix\url{https://link.aps.org/doi/10.1103/PhysRevD.96.084004}.

\bibitem[{\citenamefont{Droz et~al.}(1999)\citenamefont{Droz, Knapp, Poisson, and Owen}}]{DrEA1999}
\bibinfo{author}{\bibfnamefont{S.}~\bibnamefont{Droz}}, \bibinfo{author}{\bibfnamefont{D.~J.} \bibnamefont{Knapp}}, \bibinfo{author}{\bibfnamefont{E.}~\bibnamefont{Poisson}}, \bibnamefont{and} \bibinfo{author}{\bibfnamefont{B.~J.} \bibnamefont{Owen}}, \bibinfo{journal}{Phys. Rev. D} \textbf{\bibinfo{volume}{59}}, \bibinfo{pages}{124016} (\bibinfo{year}{1999}), \urlprefix\url{https://link.aps.org/doi/10.1103/PhysRevD.59.124016}.

\bibitem[{\citenamefont{Harms et~al.}(2013)\citenamefont{Harms, Slagmolen, Adhikari, Miller, Evans, Chen, M\"uller, and Ando}}]{HaEA2013}
\bibinfo{author}{\bibfnamefont{J.}~\bibnamefont{Harms}}, \bibinfo{author}{\bibfnamefont{B.~J.~J.} \bibnamefont{Slagmolen}}, \bibinfo{author}{\bibfnamefont{R.~X.} \bibnamefont{Adhikari}}, \bibinfo{author}{\bibfnamefont{M.~C.} \bibnamefont{Miller}}, \bibinfo{author}{\bibfnamefont{M.}~\bibnamefont{Evans}}, \bibinfo{author}{\bibfnamefont{Y.}~\bibnamefont{Chen}}, \bibinfo{author}{\bibfnamefont{H.}~\bibnamefont{M\"uller}}, \bibnamefont{and} \bibinfo{author}{\bibfnamefont{M.}~\bibnamefont{Ando}}, \bibinfo{journal}{Phys. Rev. D} \textbf{\bibinfo{volume}{88}}, \bibinfo{pages}{122003} (\bibinfo{year}{2013}), \urlprefix\url{http://link.aps.org/doi/10.1103/PhysRevD.88.122003}.

\bibitem[{\citenamefont{Sathyaprakash and Schutz}(2009)}]{SaSc2009}
\bibinfo{author}{\bibfnamefont{B.~S.} \bibnamefont{Sathyaprakash}} \bibnamefont{and} \bibinfo{author}{\bibfnamefont{B.~F.} \bibnamefont{Schutz}}, \bibinfo{journal}{Living Reviews in Relativity} \textbf{\bibinfo{volume}{12}}, \bibinfo{pages}{2} (\bibinfo{year}{2009}), ISSN \bibinfo{issn}{1433-8351}, \urlprefix\url{https://doi.org/10.12942/lrr-2009-2}.

\bibitem[{\citenamefont{G\"ursel and Tinto}(1989)}]{GuTi1989}
\bibinfo{author}{\bibfnamefont{Y.}~\bibnamefont{G\"ursel}} \bibnamefont{and} \bibinfo{author}{\bibfnamefont{M.}~\bibnamefont{Tinto}}, \bibinfo{journal}{Phys. Rev. D} \textbf{\bibinfo{volume}{40}}, \bibinfo{pages}{3884} (\bibinfo{year}{1989}), \urlprefix\url{https://link.aps.org/doi/10.1103/PhysRevD.40.3884}.

\bibitem[{\citenamefont{Regimbau et~al.}(2012)\citenamefont{Regimbau, Dent, Del~Pozzo, Giampanis, Li, Robinson, Van Den~Broeck, Meacher, Rodriguez, Sathyaprakash et~al.}}]{ReEA2012}
\bibinfo{author}{\bibfnamefont{T.}~\bibnamefont{Regimbau}}, \bibinfo{author}{\bibfnamefont{T.}~\bibnamefont{Dent}}, \bibinfo{author}{\bibfnamefont{W.}~\bibnamefont{Del~Pozzo}}, \bibinfo{author}{\bibfnamefont{S.}~\bibnamefont{Giampanis}}, \bibinfo{author}{\bibfnamefont{T.~G.~F.} \bibnamefont{Li}}, \bibinfo{author}{\bibfnamefont{C.}~\bibnamefont{Robinson}}, \bibinfo{author}{\bibfnamefont{C.}~\bibnamefont{Van Den~Broeck}}, \bibinfo{author}{\bibfnamefont{D.}~\bibnamefont{Meacher}}, \bibinfo{author}{\bibfnamefont{C.}~\bibnamefont{Rodriguez}}, \bibinfo{author}{\bibfnamefont{B.~S.} \bibnamefont{Sathyaprakash}}, \bibnamefont{et~al.}, \bibinfo{journal}{Phys. Rev. D} \textbf{\bibinfo{volume}{86}}, \bibinfo{pages}{122001} (\bibinfo{year}{2012}), \urlprefix\url{https://link.aps.org/doi/10.1103/PhysRevD.86.122001}.

\bibitem[{\citenamefont{Parida et~al.}(2016)\citenamefont{Parida, Mitra, and Jhingan}}]{PMJ2016}
\bibinfo{author}{\bibfnamefont{A.}~\bibnamefont{Parida}}, \bibinfo{author}{\bibfnamefont{S.}~\bibnamefont{Mitra}}, \bibnamefont{and} \bibinfo{author}{\bibfnamefont{S.}~\bibnamefont{Jhingan}}, \bibinfo{journal}{Journal of Cosmology and Astroparticle Physics} \textbf{\bibinfo{volume}{2016}}, \bibinfo{pages}{024} (\bibinfo{year}{2016}), \urlprefix\url{https://doi.org/10.1088/1475-7516/2016/04/024}.

\bibitem[{\citenamefont{Press et~al.}(2007)\citenamefont{Press, Teukolsky, Vetterling, and Flannery}}]{PrEA2007}
\bibinfo{author}{\bibfnamefont{W.~H.} \bibnamefont{Press}}, \bibinfo{author}{\bibfnamefont{S.~A.} \bibnamefont{Teukolsky}}, \bibinfo{author}{\bibfnamefont{W.~T.} \bibnamefont{Vetterling}}, \bibnamefont{and} \bibinfo{author}{\bibfnamefont{B.~P.} \bibnamefont{Flannery}}, \emph{\bibinfo{title}{{Numerical recipes 3rd edition: The art of scientific computing}}} (\bibinfo{publisher}{Cambridge University Press}, \bibinfo{year}{2007}).

\bibitem[{\citenamefont{{LIGO Scientific Collaboration}}(2018)}]{lalsuite}
\bibinfo{author}{\bibnamefont{{LIGO Scientific Collaboration}}}, \emph{\bibinfo{title}{{LIGO} {A}lgorithm {L}ibrary - {LALS}uite}}, \bibinfo{howpublished}{free software (GPL)} (\bibinfo{year}{2018}).

\bibitem[{\citenamefont{{Ng} et~al.}(2021)\citenamefont{{Ng}, {Chen}, {Goncharov}, {Dupletsa}, {Borhanian}, {Branchesi}, {Harms}, {Maggiore}, {Sathyaprakash}, and {Vitale}}}]{Ng2021}
\bibinfo{author}{\bibfnamefont{K.~K.~Y.} \bibnamefont{{Ng}}}, \bibinfo{author}{\bibfnamefont{S.}~\bibnamefont{{Chen}}}, \bibinfo{author}{\bibfnamefont{B.}~\bibnamefont{{Goncharov}}}, \bibinfo{author}{\bibfnamefont{U.}~\bibnamefont{{Dupletsa}}}, \bibinfo{author}{\bibfnamefont{S.}~\bibnamefont{{Borhanian}}}, \bibinfo{author}{\bibfnamefont{M.}~\bibnamefont{{Branchesi}}}, \bibinfo{author}{\bibfnamefont{J.}~\bibnamefont{{Harms}}}, \bibinfo{author}{\bibfnamefont{M.}~\bibnamefont{{Maggiore}}}, \bibinfo{author}{\bibfnamefont{B.~S.} \bibnamefont{{Sathyaprakash}}}, \bibnamefont{and} \bibinfo{author}{\bibfnamefont{S.}~\bibnamefont{{Vitale}}}, \bibinfo{journal}{arXiv e-prints} \bibinfo{eid}{arXiv:2108.07276} (\bibinfo{year}{2021}), \eprint{2108.07276}.

\bibitem[{\citenamefont{Luca et~al.}(2021)\citenamefont{Luca, Franciolini, Pani, and Riotto}}]{DeLuca2021}
\bibinfo{author}{\bibfnamefont{V.~D.} \bibnamefont{Luca}}, \bibinfo{author}{\bibfnamefont{G.}~\bibnamefont{Franciolini}}, \bibinfo{author}{\bibfnamefont{P.}~\bibnamefont{Pani}}, \bibnamefont{and} \bibinfo{author}{\bibfnamefont{A.}~\bibnamefont{Riotto}}, \bibinfo{journal}{Journal of Cosmology and Astroparticle Physics} \textbf{\bibinfo{volume}{2021}}, \bibinfo{pages}{003} (\bibinfo{year}{2021}), \urlprefix\url{https://doi.org/10.1088/1475-7516/2021/05/003}.

\bibitem[{\citenamefont{Ashton et~al.}(2019)\citenamefont{Ashton, Hübner, Lasky, Talbot, Ackley, Biscoveanu, Chu, Divakarla, Easter, Goncharov et~al.}}]{AsEA2019}
\bibinfo{author}{\bibfnamefont{G.}~\bibnamefont{Ashton}}, \bibinfo{author}{\bibfnamefont{M.}~\bibnamefont{Hübner}}, \bibinfo{author}{\bibfnamefont{P.~D.} \bibnamefont{Lasky}}, \bibinfo{author}{\bibfnamefont{C.}~\bibnamefont{Talbot}}, \bibinfo{author}{\bibfnamefont{K.}~\bibnamefont{Ackley}}, \bibinfo{author}{\bibfnamefont{S.}~\bibnamefont{Biscoveanu}}, \bibinfo{author}{\bibfnamefont{Q.}~\bibnamefont{Chu}}, \bibinfo{author}{\bibfnamefont{A.}~\bibnamefont{Divakarla}}, \bibinfo{author}{\bibfnamefont{P.~J.} \bibnamefont{Easter}}, \bibinfo{author}{\bibfnamefont{B.}~\bibnamefont{Goncharov}}, \bibnamefont{et~al.}, \bibinfo{journal}{The Astrophysical Journal Supplement Series} \textbf{\bibinfo{volume}{241}}, \bibinfo{pages}{27} (\bibinfo{year}{2019}), \urlprefix\url{https://doi.org/10.3847%2F1538-4365%2Fab06fc}.

\bibitem[{\citenamefont{Vallisneri}(2011)}]{Val2011}
\bibinfo{author}{\bibfnamefont{M.}~\bibnamefont{Vallisneri}}, \bibinfo{journal}{Phys. Rev. Lett.} \textbf{\bibinfo{volume}{107}}, \bibinfo{pages}{191104} (\bibinfo{year}{2011}), \urlprefix\url{https://link.aps.org/doi/10.1103/PhysRevLett.107.191104}.

\bibitem[{\citenamefont{Hoy and Raymond}(2021)}]{Hoy:2020vys}
\bibinfo{author}{\bibfnamefont{C.}~\bibnamefont{Hoy}} \bibnamefont{and} \bibinfo{author}{\bibfnamefont{V.}~\bibnamefont{Raymond}}, \bibinfo{journal}{SoftwareX} \textbf{\bibinfo{volume}{15}}, \bibinfo{pages}{100765} (\bibinfo{year}{2021}), \eprint{2006.06639}.

\bibitem[{\citenamefont{Buonanno et~al.}(2009)\citenamefont{Buonanno, Iyer, Ochsner, Pan, and Sathyaprakash}}]{TF2_1}
\bibinfo{author}{\bibfnamefont{A.}~\bibnamefont{Buonanno}}, \bibinfo{author}{\bibfnamefont{B.~R.} \bibnamefont{Iyer}}, \bibinfo{author}{\bibfnamefont{E.}~\bibnamefont{Ochsner}}, \bibinfo{author}{\bibfnamefont{Y.}~\bibnamefont{Pan}}, \bibnamefont{and} \bibinfo{author}{\bibfnamefont{B.~S.} \bibnamefont{Sathyaprakash}}, \bibinfo{journal}{Physical Review D} \textbf{\bibinfo{volume}{80}} (\bibinfo{year}{2009}), ISSN \bibinfo{issn}{1550-2368}, \urlprefix\url{http://dx.doi.org/10.1103/PhysRevD.80.084043}.

\bibitem[{\citenamefont{Isoyama et~al.}(2021)\citenamefont{Isoyama, Sturani, and Nakano}}]{TF2_2}
\bibinfo{author}{\bibfnamefont{S.}~\bibnamefont{Isoyama}}, \bibinfo{author}{\bibfnamefont{R.}~\bibnamefont{Sturani}}, \bibnamefont{and} \bibinfo{author}{\bibfnamefont{H.}~\bibnamefont{Nakano}}, \bibinfo{journal}{Handbook of Gravitational Wave Astronomy} p. \bibinfo{pages}{1–49} (\bibinfo{year}{2021}), \urlprefix\url{http://dx.doi.org/10.1007/978-981-15-4702-7_31-1}.

\bibitem[{\citenamefont{Husa et~al.}(2016)\citenamefont{Husa, Khan, Hannam, Pürrer, Ohme, Forteza, and Bohé}}]{Phenom1}
\bibinfo{author}{\bibfnamefont{S.}~\bibnamefont{Husa}}, \bibinfo{author}{\bibfnamefont{S.}~\bibnamefont{Khan}}, \bibinfo{author}{\bibfnamefont{M.}~\bibnamefont{Hannam}}, \bibinfo{author}{\bibfnamefont{M.}~\bibnamefont{Pürrer}}, \bibinfo{author}{\bibfnamefont{F.}~\bibnamefont{Ohme}}, \bibinfo{author}{\bibfnamefont{X.~J.} \bibnamefont{Forteza}}, \bibnamefont{and} \bibinfo{author}{\bibfnamefont{A.}~\bibnamefont{Bohé}}, \bibinfo{journal}{Physical Review D} \textbf{\bibinfo{volume}{93}} (\bibinfo{year}{2016}), ISSN \bibinfo{issn}{2470-0029}, \urlprefix\url{http://dx.doi.org/10.1103/PhysRevD.93.044006}.

\bibitem[{\citenamefont{Khan et~al.}(2016)\citenamefont{Khan, Husa, Hannam, Ohme, Pürrer, Forteza, and Bohé}}]{Phenom2}
\bibinfo{author}{\bibfnamefont{S.}~\bibnamefont{Khan}}, \bibinfo{author}{\bibfnamefont{S.}~\bibnamefont{Husa}}, \bibinfo{author}{\bibfnamefont{M.}~\bibnamefont{Hannam}}, \bibinfo{author}{\bibfnamefont{F.}~\bibnamefont{Ohme}}, \bibinfo{author}{\bibfnamefont{M.}~\bibnamefont{Pürrer}}, \bibinfo{author}{\bibfnamefont{X.~J.} \bibnamefont{Forteza}}, \bibnamefont{and} \bibinfo{author}{\bibfnamefont{A.}~\bibnamefont{Bohé}}, \bibinfo{journal}{Physical Review D} \textbf{\bibinfo{volume}{93}} (\bibinfo{year}{2016}), ISSN \bibinfo{issn}{2470-0029}, \urlprefix\url{http://dx.doi.org/10.1103/PhysRevD.93.044007}.

\bibitem[{\citenamefont{Broeck and Sengupta}(2007)}]{BrSe2007}
\bibinfo{author}{\bibfnamefont{C.~V.~D.} \bibnamefont{Broeck}} \bibnamefont{and} \bibinfo{author}{\bibfnamefont{A.~S.} \bibnamefont{Sengupta}}, \bibinfo{journal}{Classical and Quantum Gravity} \textbf{\bibinfo{volume}{24}}, \bibinfo{pages}{1089} (\bibinfo{year}{2007}), \urlprefix\url{https://doi.org/10.1088%2F0264-9381%2F24%2F5%2F005}.

\bibitem[{\citenamefont{Khan et~al.}(2020)\citenamefont{Khan, Ohme, Chatziioannou, and Hannam}}]{KhEA2020}
\bibinfo{author}{\bibfnamefont{S.}~\bibnamefont{Khan}}, \bibinfo{author}{\bibfnamefont{F.}~\bibnamefont{Ohme}}, \bibinfo{author}{\bibfnamefont{K.}~\bibnamefont{Chatziioannou}}, \bibnamefont{and} \bibinfo{author}{\bibfnamefont{M.}~\bibnamefont{Hannam}}, \bibinfo{journal}{Phys. Rev. D} \textbf{\bibinfo{volume}{101}}, \bibinfo{pages}{024056} (\bibinfo{year}{2020}), \urlprefix\url{https://link.aps.org/doi/10.1103/PhysRevD.101.024056}.

\bibitem[{\citenamefont{Usman et~al.}(2019)\citenamefont{Usman, Mills, and Fairhurst}}]{UMF2019}
\bibinfo{author}{\bibfnamefont{S.~A.} \bibnamefont{Usman}}, \bibinfo{author}{\bibfnamefont{J.~C.} \bibnamefont{Mills}}, \bibnamefont{and} \bibinfo{author}{\bibfnamefont{S.}~\bibnamefont{Fairhurst}}, \bibinfo{journal}{The Astrophysical Journal} \textbf{\bibinfo{volume}{877}}, \bibinfo{pages}{82} (\bibinfo{year}{2019}), \urlprefix\url{https://doi.org/10.3847/1538-4357/ab0b3e}.

\bibitem[{\citenamefont{Chen et~al.}(2019)\citenamefont{Chen, Vitale, and Narayan}}]{CVN2019}
\bibinfo{author}{\bibfnamefont{H.-Y.} \bibnamefont{Chen}}, \bibinfo{author}{\bibfnamefont{S.}~\bibnamefont{Vitale}}, \bibnamefont{and} \bibinfo{author}{\bibfnamefont{R.}~\bibnamefont{Narayan}}, \bibinfo{journal}{Phys. Rev. X} \textbf{\bibinfo{volume}{9}}, \bibinfo{pages}{031028} (\bibinfo{year}{2019}), \urlprefix\url{https://link.aps.org/doi/10.1103/PhysRevX.9.031028}.

\bibitem[{\citenamefont{Wen and Chen}(2010)}]{WeCh2010}
\bibinfo{author}{\bibfnamefont{L.}~\bibnamefont{Wen}} \bibnamefont{and} \bibinfo{author}{\bibfnamefont{Y.}~\bibnamefont{Chen}}, \bibinfo{journal}{Phys. Rev. D} \textbf{\bibinfo{volume}{81}}, \bibinfo{pages}{082001} (\bibinfo{year}{2010}), \urlprefix\url{https://link.aps.org/doi/10.1103/PhysRevD.81.082001}.

\bibitem[{\citenamefont{Li et~al.}(2022)\citenamefont{Li, Heng, Chan, Messenger, and Fan}}]{Li_2022}
\bibinfo{author}{\bibfnamefont{Y.}~\bibnamefont{Li}}, \bibinfo{author}{\bibfnamefont{I.~S.} \bibnamefont{Heng}}, \bibinfo{author}{\bibfnamefont{M.~L.} \bibnamefont{Chan}}, \bibinfo{author}{\bibfnamefont{C.}~\bibnamefont{Messenger}}, \bibnamefont{and} \bibinfo{author}{\bibfnamefont{X.}~\bibnamefont{Fan}}, \bibinfo{journal}{Physical Review D} \textbf{\bibinfo{volume}{105}} (\bibinfo{year}{2022}), \urlprefix\url{https://doi.org/10.1103%2Fphysrevd.105.043010}.

\bibitem[{\citenamefont{Hild et~al.}(2011)\citenamefont{Hild, Abernathy, Acernese, Amaro-Seoane, Andersson, Arun, Barone, Barr, Barsuglia, Beker et~al.}}]{HiEA2011}
\bibinfo{author}{\bibfnamefont{S.}~\bibnamefont{Hild}}, \bibinfo{author}{\bibfnamefont{M.}~\bibnamefont{Abernathy}}, \bibinfo{author}{\bibfnamefont{F.}~\bibnamefont{Acernese}}, \bibinfo{author}{\bibfnamefont{P.}~\bibnamefont{Amaro-Seoane}}, \bibinfo{author}{\bibfnamefont{N.}~\bibnamefont{Andersson}}, \bibinfo{author}{\bibfnamefont{K.}~\bibnamefont{Arun}}, \bibinfo{author}{\bibfnamefont{F.}~\bibnamefont{Barone}}, \bibinfo{author}{\bibfnamefont{B.}~\bibnamefont{Barr}}, \bibinfo{author}{\bibfnamefont{M.}~\bibnamefont{Barsuglia}}, \bibinfo{author}{\bibfnamefont{M.}~\bibnamefont{Beker}}, \bibnamefont{et~al.}, \bibinfo{journal}{Classical and Quantum Gravity} \textbf{\bibinfo{volume}{28}}, \bibinfo{pages}{094013} (\bibinfo{year}{2011}), \urlprefix\url{http://stacks.iop.org/0264-9381/28/i=9/a=094013}.

\bibitem[{\citenamefont{Abbott et~al.}(2019)\citenamefont{Abbott, Abbott, Abbott, Abraham, Acernese, Ackley, Adams, Adhikari, Adya, Affeldt et~al.}}]{AbEA2019c}
\bibinfo{author}{\bibfnamefont{B.~P.} \bibnamefont{Abbott}}, \bibinfo{author}{\bibfnamefont{R.}~\bibnamefont{Abbott}}, \bibinfo{author}{\bibfnamefont{T.~D.} \bibnamefont{Abbott}}, \bibinfo{author}{\bibfnamefont{S.}~\bibnamefont{Abraham}}, \bibinfo{author}{\bibfnamefont{F.}~\bibnamefont{Acernese}}, \bibinfo{author}{\bibfnamefont{K.}~\bibnamefont{Ackley}}, \bibinfo{author}{\bibfnamefont{C.}~\bibnamefont{Adams}}, \bibinfo{author}{\bibfnamefont{R.~X.} \bibnamefont{Adhikari}}, \bibinfo{author}{\bibfnamefont{V.~B.} \bibnamefont{Adya}}, \bibinfo{author}{\bibfnamefont{C.}~\bibnamefont{Affeldt}}, \bibnamefont{et~al.} (\bibinfo{collaboration}{LIGO Scientific Collaboration and Virgo Collaboration}), \bibinfo{journal}{Phys. Rev. X} \textbf{\bibinfo{volume}{9}}, \bibinfo{pages}{031040} (\bibinfo{year}{2019}), \urlprefix\url{https://link.aps.org/doi/10.1103/PhysRevX.9.031040}.

\bibitem[{\citenamefont{Abbott et~al.}(2021)\citenamefont{Abbott, Abbott, Abraham, Acernese, Ackley, Adams, Adams, Adhikari, Adya, Affeldt et~al.}}]{AbEA2021}
\bibinfo{author}{\bibfnamefont{R.}~\bibnamefont{Abbott}}, \bibinfo{author}{\bibfnamefont{T.~D.} \bibnamefont{Abbott}}, \bibinfo{author}{\bibfnamefont{S.}~\bibnamefont{Abraham}}, \bibinfo{author}{\bibfnamefont{F.}~\bibnamefont{Acernese}}, \bibinfo{author}{\bibfnamefont{K.}~\bibnamefont{Ackley}}, \bibinfo{author}{\bibfnamefont{A.}~\bibnamefont{Adams}}, \bibinfo{author}{\bibfnamefont{C.}~\bibnamefont{Adams}}, \bibinfo{author}{\bibfnamefont{R.~X.} \bibnamefont{Adhikari}}, \bibinfo{author}{\bibfnamefont{V.~B.} \bibnamefont{Adya}}, \bibinfo{author}{\bibfnamefont{C.}~\bibnamefont{Affeldt}}, \bibnamefont{et~al.} (\bibinfo{collaboration}{LIGO Scientific Collaboration and Virgo Collaboration}), \bibinfo{journal}{Phys. Rev. X} \textbf{\bibinfo{volume}{11}}, \bibinfo{pages}{021053} (\bibinfo{year}{2021}), \urlprefix\url{https://link.aps.org/doi/10.1103/PhysRevX.11.021053}.

\bibitem[{\citenamefont{{The LIGO Scientific Collaboration} et~al.}(2021)\citenamefont{{The LIGO Scientific Collaboration}, {the Virgo Collaboration}, {the KAGRA Collaboration}, Abbott, Abbott, Acernese, Ackley, Adams, Adhikari, Adhikari et~al.}}]{AbEA2021a}
\bibinfo{author}{\bibnamefont{{The LIGO Scientific Collaboration}}}, \bibinfo{author}{\bibnamefont{{the Virgo Collaboration}}}, \bibinfo{author}{\bibnamefont{{the KAGRA Collaboration}}}, \bibinfo{author}{\bibfnamefont{R.}~\bibnamefont{Abbott}}, \bibinfo{author}{\bibfnamefont{T.~D.} \bibnamefont{Abbott}}, \bibinfo{author}{\bibfnamefont{F.}~\bibnamefont{Acernese}}, \bibinfo{author}{\bibfnamefont{K.}~\bibnamefont{Ackley}}, \bibinfo{author}{\bibfnamefont{C.}~\bibnamefont{Adams}}, \bibinfo{author}{\bibfnamefont{N.}~\bibnamefont{Adhikari}}, \bibinfo{author}{\bibfnamefont{R.~X.} \bibnamefont{Adhikari}}, \bibnamefont{et~al.}, \emph{\bibinfo{title}{{GWTC-3: Compact Binary Coalescences Observed by LIGO and Virgo During the Second Part of the Third Observing Run}}} (\bibinfo{year}{2021}), \urlprefix\url{https://arxiv.org/abs/2111.03606}.

\bibitem[{\citenamefont{Wette}(2020)}]{Wet2020}
\bibinfo{author}{\bibfnamefont{K.}~\bibnamefont{Wette}}, \bibinfo{journal}{SoftwareX} \textbf{\bibinfo{volume}{12}}, \bibinfo{pages}{100634} (\bibinfo{year}{2020}), ISSN \bibinfo{issn}{2352-7110}, \urlprefix\url{https://www.sciencedirect.com/science/article/pii/S2352711020303472}.

\end{thebibliography}

\appendix

\section{Posterior samples}

\begin{figure*}[ht!]
\includegraphics[width=0.9\textwidth]{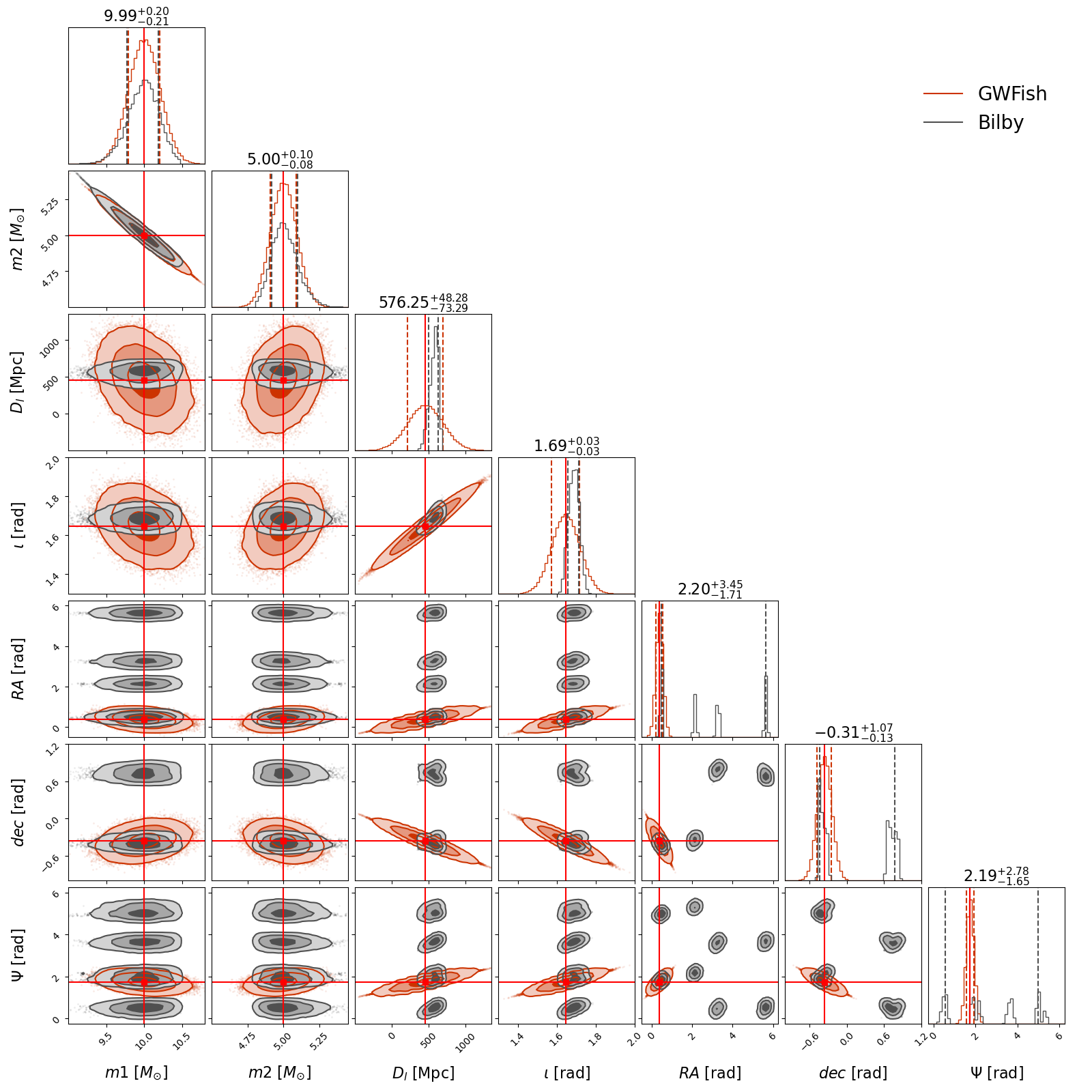}
\caption{Posterior sampling output for Signal 1 ($5M_{\odot}+10M_{\odot}$).}
\label{fig:signal1_corner}
\end{figure*}

\begin{figure*}[ht!]
\includegraphics[width=0.9\textwidth]{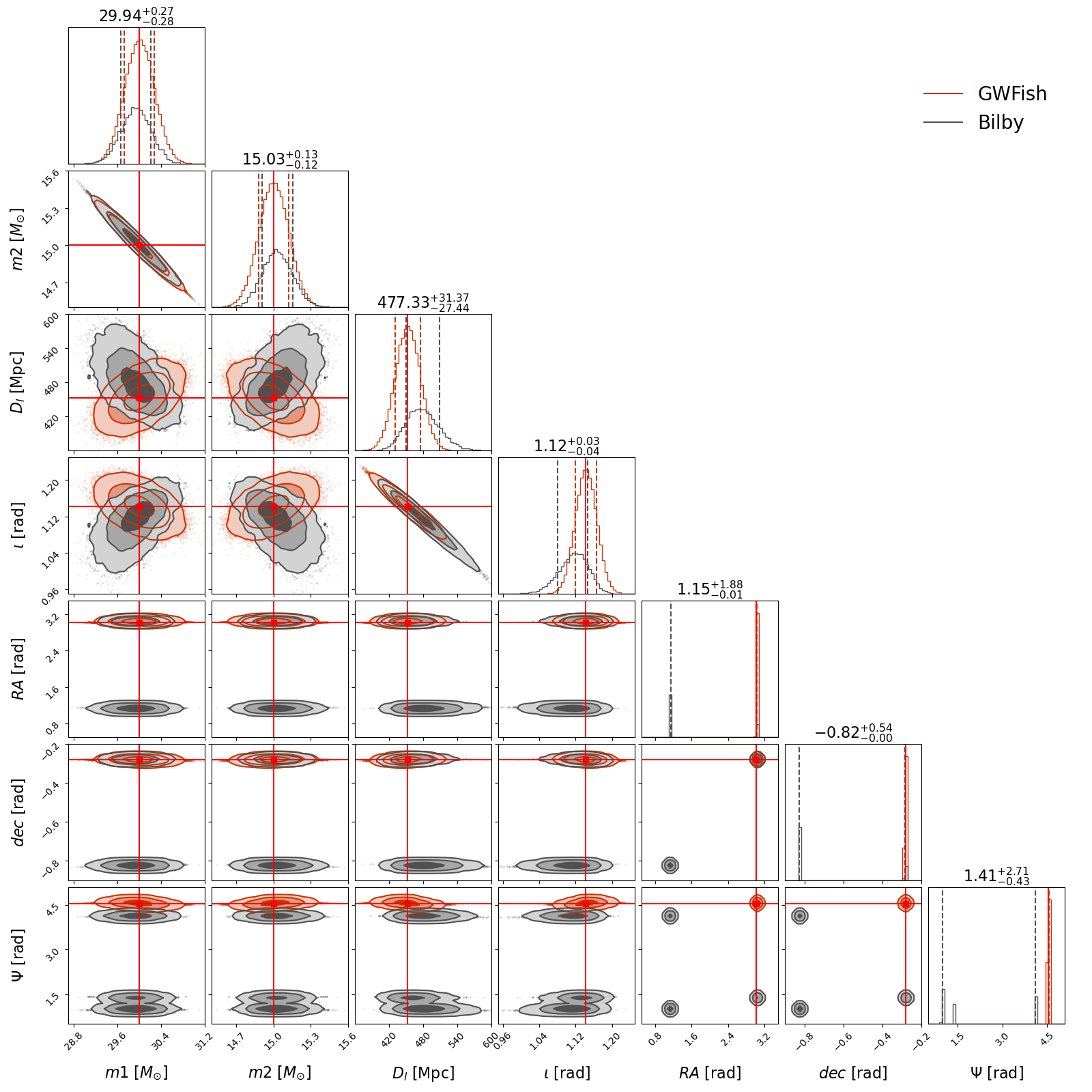}
\caption{Posterior sampling output for Signal 2 ($15M_{\odot}+30M_{\odot}$).}
\label{fig:signal2_corner}
\end{figure*}

\begin{figure*}[ht!]
\includegraphics[width=0.9\textwidth]{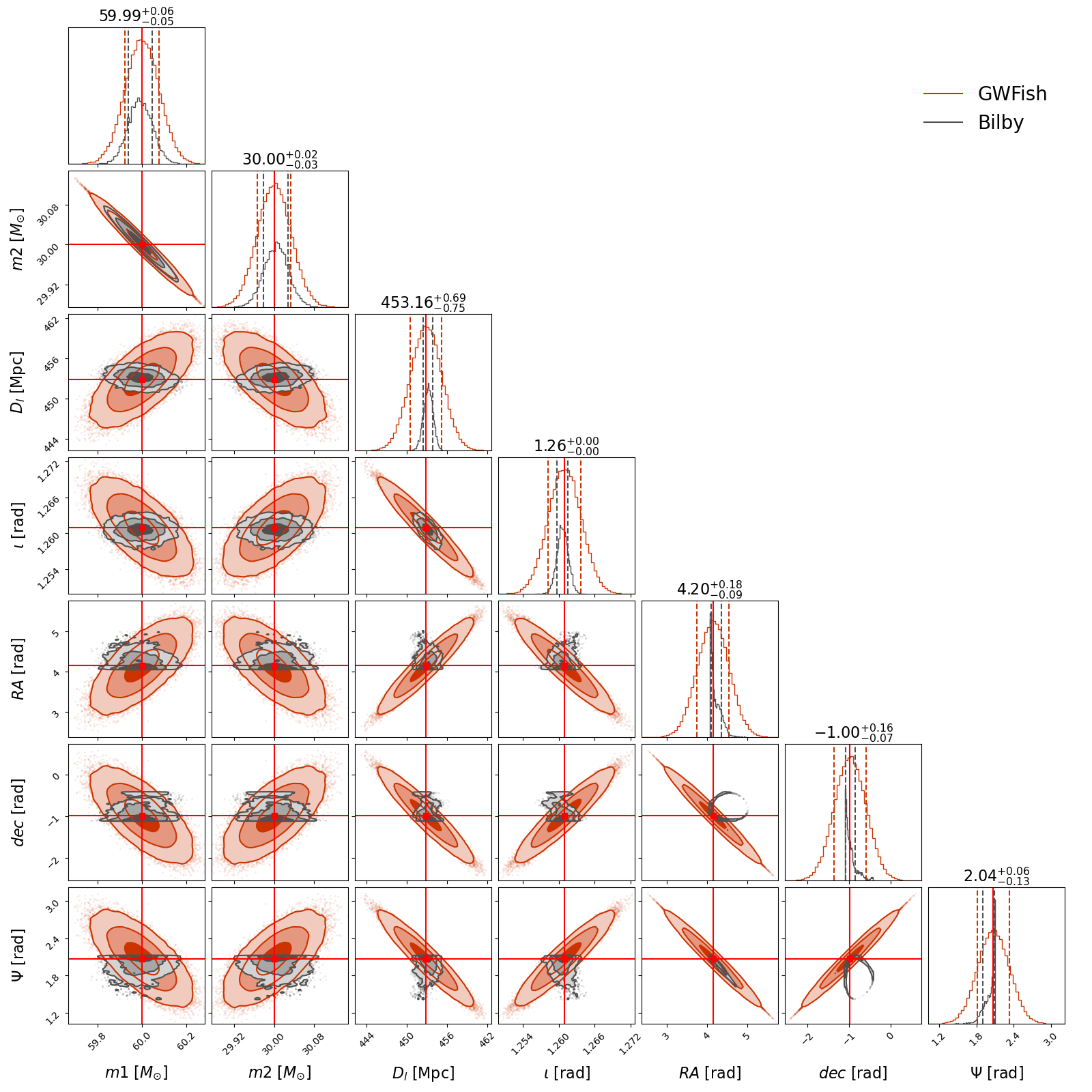}
\caption{Posterior sampling output for Signal 3 ($30M_{\odot}+60M_{\odot}$).}
\label{fig:signal3_corner}
\end{figure*}

\begin{figure*}[ht!]
\includegraphics[width=0.9\textwidth]{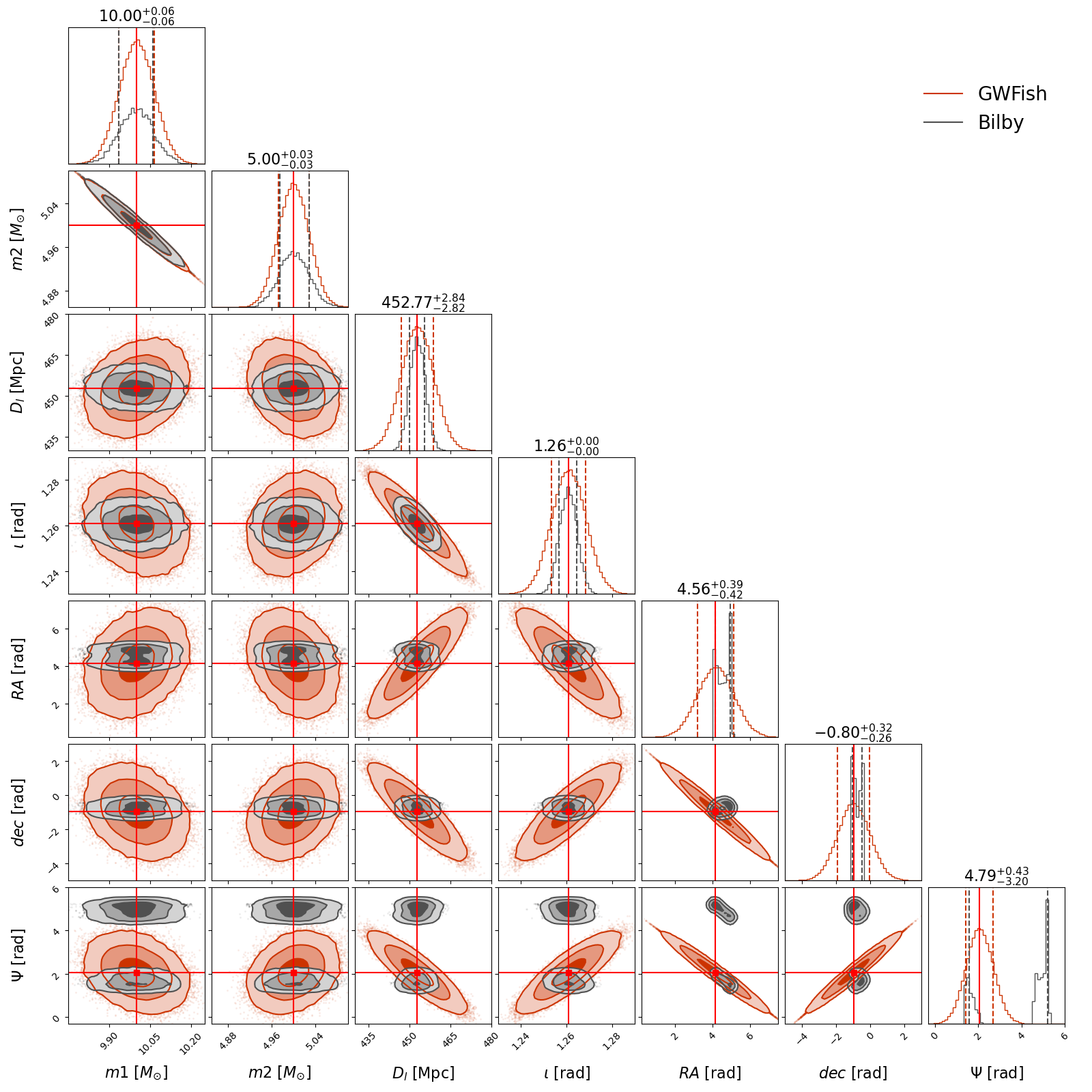}
\caption{Posterior sampling output for Signal 3* ($5M_{\odot}+10M_{\odot}$).}
\label{fig:signal4_corner}
\end{figure*}

\begin{figure*}[ht!]
\includegraphics[width=0.9\textwidth]{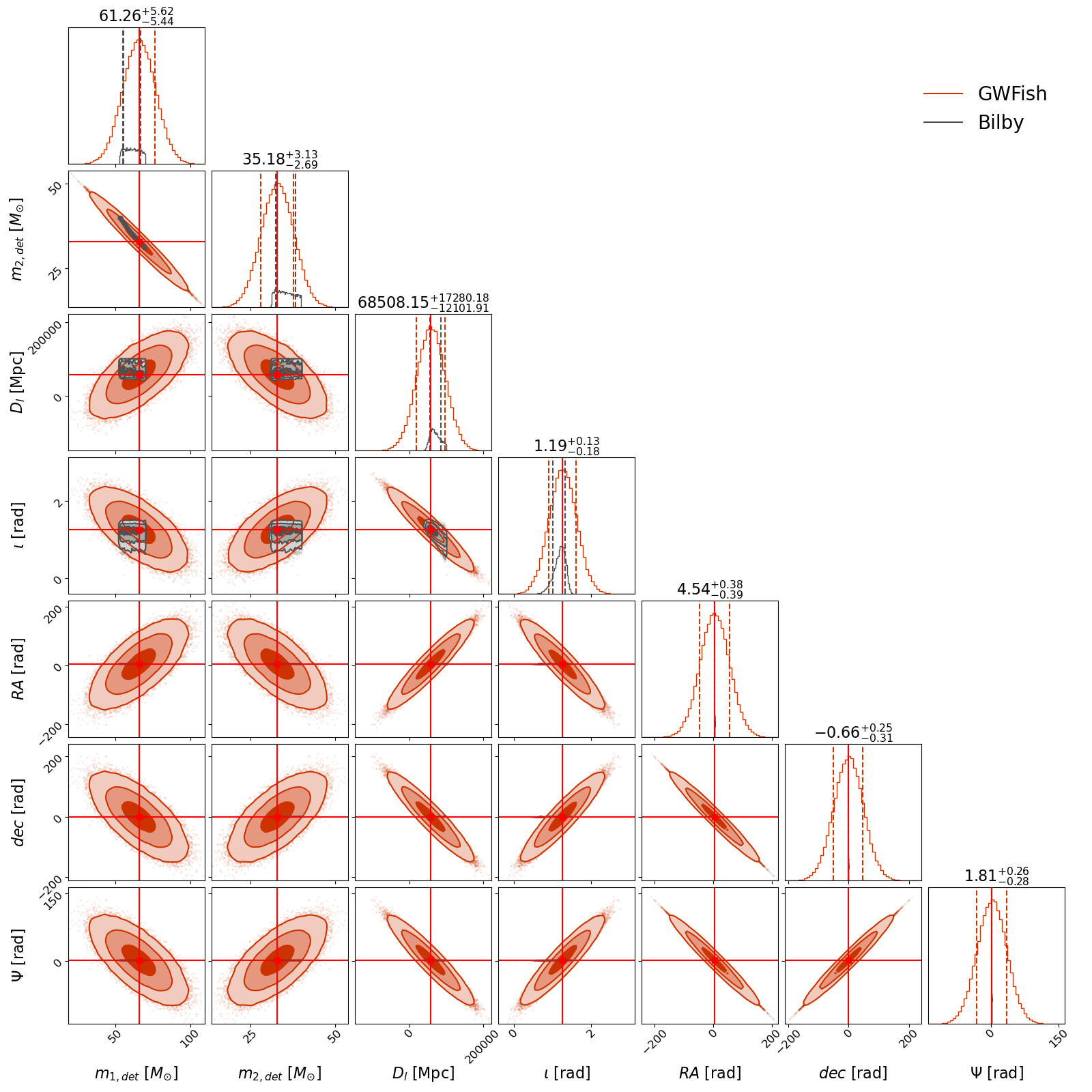}
\caption{Posterior sampling output for Signal 3 with increased distance so that SNR=8. Especially for the sky location and polarization angle, the posterior obtained from \textsc{bilby} is much narrower than the \textsc{gwfish} errors and invisible in the corner plot.}
\label{fig:signal3_low_snr_corner}
\end{figure*}

Figures \ref{fig:signal1_corner}, \ref{fig:signal2_corner}, \ref{fig:signal3_corner} and \ref{fig:signal4_corner} show corner plots of the posteriors obtained with \textsc{bilby} for signals 1, 2, 3 and 3* respectively, whose parameters are described in table \ref{tab:signal_sum}. Superimposed we report the results from \textsc{gwfish}. The last corner plot in Figure \ref{fig:signal3_low_snr_corner} shows a low-SNR example. The \textsc{bilby} PE errors of all angular parameters except for the inclination angle are much smaller than the \textsc{gwfish} error estimates, which causes the \textsc{bilby} marginal distributions to be hardly visible in these cases.

\end{document}